\documentclass[journal]{IEEEtran}

\ifCLASSINFOpdf
\else
\fi
\usepackage{epsfig}
\usepackage{graphicx}
\usepackage{amsbsy}
\usepackage{amsmath}
\usepackage{amssymb}
\usepackage{euscript}

\newtheorem{mytheorem}{\bf Theorem}
\newtheorem{mylemma}{\bf Lemma}

\newtheorem{corollary}{\bf Corollary}

\newtheorem{myremark}{\textbf{Remark}}

\newcommand {\Define} {\stackrel {\Delta} {=}  }

\begin{document}
\title{Impact of Transceiver Power Consumption on the Energy Efficiency of Zero-Forcing Detector in Massive MIMO Systems}
\author{\IEEEauthorblockN{Saif Khan Mohammed}
\thanks{Manuscript received Jan. 20, 2014; revised June 15, 2014 and Oct. 7, 2014; accepted Oct. 13, 2014.
The editor coordinating the review of this paper and approving it for publication
was Prof. Tony Q. S. Quek.}
\thanks{
Saif Khan Mohammed is currently with the Dept. of Electrical Engineering, Indian Institute of Technology (I.I.T.), Delhi, India. He is also associated with the Bharti School of Telecommunication Technology and Management (BSTTM), I.I.T. Delhi. This work is supported by EMR funding from the Science and Engineering Research Board (SERB), Department of Science and Technology, Government of India.
}
}

\markboth{IEEE Transactions on Communications,~Vol.~xx, No.~xx, xx~xxxx}
        {Mohammed, S. K. : Impact of Transceiver Power Consumption on the Energy Efficiency of Zero-Forcing Detector in Massive MIMO Systems}

\maketitle

\begin{abstract}
We consider the impact of transceiver power consumption on the energy efficiency (EE) of the Zero Forcing (ZF) detector in the uplink of massive MIMO systems, where a base station (BS) with $M$ antennas
communicates coherently with $K$ single antenna user terminals (UTs). We consider the problem of maximizing the EE with respect to $(M,K)$ for a fixed sum spectral efficiency. Through analysis we study the impact of system parameters on the optimal EE. System parameters consists of the average channel gain to the users and the power consumption parameters (PCPs) (e.g., power consumed by each RF antenna/receiver at BS). When the average user channel gain is high or else the BS/UT design is power inefficient, our analysis reveals that it is optimal to have a few BS antennas and a single user, i.e., non-massive MIMO regime. Similarly, when the channel gain is small or else the BS/UT design is power efficient, it is optimal of have a larger $(M,K)$, i.e., massive MIMO regime. Tight analytical bounds on the optimal EE are proposed for both these regimes.
The impact of the system parameters on the optimal EE is studied
and several interesting insights are drawn.
\end{abstract}

\section{Introduction}\label{sec:SysModel}
In recent years there has been a surge of interest on energy efficient ``green communication'' systems,
primarily arising out of environmental and cost concerns due to the every increasing power consumption of cellular systems \cite{Zanders}. 
Fifth generation cellular communication systems (5G) are expected to significantly improve the total system capacity as well as energy efficiency compared to 4G systems \cite{Whatis5G}.
Most of this improvement is expected to be achieved through, i) network densification (i.e., more base station nodes per unit area), ii) increased system bandwidth (e.g. usage of mmWave spectrum) and iii) massive MIMO \cite{Whatis5G, smallcells}.
In this paper, we are interested in studying the energy efficiency of massive MIMO based cellular systems. 
Massive MIMO Systems/ Large MIMO Systems/ Large Scale Antenna Systems collectively refer to a communication system where a base station (BS) with $M$ antennas (several tens to hundred) communicates coherently with $K$ users (few tens) on the same time-frequency resource \cite{SPM-paper}, \cite{TM}, \cite{MM}.
Large antenna arrays at the BS helps in realizing beamforming gains which significantly improves the energy efficiency (EE). The
spectral efficiency energy efficiency tradeoff of massive MIMO systems has recently been studied in \cite{HNQ}.
However, in \cite{HNQ} only the power consumed by the power amplifiers (PA) at the user terminals (UT) has been considered. With large $M$ the total power consumed by the $M$ RF receivers
at the BS becomes significant and must therefore be taken into consideration \cite{Bjorn1}.
In \cite{Bjorn1} the EE of massive MIMO systems is maximized w.r.t. $M$ but the impact of power consumption parameters (e.g., per antenna power consumption at BS/UT) on the optimal $M$ and the optimal EE
has not been studied analytically.  

The impact of transceiver power consumption on the EE of MIMO systems has been recently considered
in \cite{Miao,DHa,HYang,Bjorn}. In \cite{Miao} it is shown that the EE of uplink MIMO systems can be optimized by selectively turning off antennas at the UT. In \cite{DHa},
the authors optimize the EE of downlink massive MIMO systems with respect to (w.r.t.) the number of BS antennas.
It is shown that the EE is a quasi-concave function of the number of BS antennas.
In \cite{HYang} downlink massive MIMO systems are considered, and for a fixed $M$ the EE is maximized
w.r.t. the total power radiated from the BS and the number of UTs. However, results in \cite{Miao,DHa,HYang} are based
on numerical simulations and therefore they provide little insight on the effect of system parameters
(e.g., cell size, power consumed by each RF receiver antenna) on the optimized EE.

In \cite{Bjorn}, the authors consider the downlink of a multiuser MIMO system, and for the ZF precoder they analytically optimize the EE separately w.r.t. $M$, $K$ and the total power radiated from the BS.
They show the very interesting result that massive MIMO must be used to increase EE only when interference suppressing multiuser precoding schemes (e.g., ZF, MMSE) are used. They however do not analytically characterize
the effect of changing system parameters on the optimal EE. Also, no analytical condition (in terms of the system parameters) has been proposed to decide as to when the system must operate in the massive MIMO regime and when not.

In this paper, we consider the uplink of a single-cell multiuser MIMO channel, where each UT is equipped
with a single antenna. For a fixed sum spectral efficiency $R$
and fixed system parameters $\Theta$, we propose to maximize the EE when a ZF receiver\footnote{\footnotesize{Among the low complexity receivers, we consider the ZF receiver due to its better
ability to cancel multi-user interference as compared to the maximum ratio combining (MRC) receiver \cite{HNQ}, specially when $R$ is large. Due to this reason, the ZF receiver is expected to have a higher EE than the MRC receiver \cite{Bjorn}. It is also known that for sufficiently large $M/K$, the MMSE and the ZF receiver have similar performance in terms of the transmit power requirement for the UTs to achieve a given fixed sum rate $R$ \cite{HNQ}. Due to similar detection complexity, the MMSE and the ZF receivers are therefore expected to have similar EE, as has been shown recently in \cite{Bjorn} assuming perfect channel knowledge.}} is used for multiuser detection at the BS. The system parameter $\Theta$ consists of the average channel gain to the users, and the power consumption parameters (PCPs) (e.g., power consumed by each RF antenna/receiver at BS, power consumed by the transmitter circuitry
at each UT, power consumed for ZF multiuser detection/channel estimation at the BS).
The system model is discussed in Section \ref{sys_model}, whereas the power consumption model and the proposed EE maximization problem is presented
in Section \ref{sec_pow_model}.
In Section \ref{subsec_largerho}, we find analytical conditions on $(R, \Theta)$ such that
the optimal EE is achieved by having very few BS antennas and a single user (i.e.,
non-massive MIMO regime). For a given $(R, \Theta)$ these conditions are met when either power inefficient
hardware design is used at the BS and UTs, or when the cell size is sufficiently small.
In the non-massive MIMO regime, the optimal EE is observed to decrease linearly with increasing
value of the PCPs.

Our analysis and simulations reveal that with fixed $R$ and increasing cell size/reduction in
the value of PCPs, the optimal $M$ starts increasing. The optimal number of users
is more than one, but is limited by the number of channel uses per coherence interval.
This is referred to as the massive MIMO regime and is studied in Section \ref{subsec_smallrho}.
We specifically consider those scenarios where the number of users is limited to a few tens, since
the channel rank is anyways limited by the amount of physical scattering.
We also consider only those channels which have a coherence interval sufficiently large compared
to the $K$ channel uses required for acquiring channel estimates.
In the massive MIMO regime, our analysis of the EE suggests that for a fixed $R$ and fixed power consumption
parameters, the optimal EE decreases with increasing cell size.
Interestingly, due to varying $(M,K)$, this decrease is found to be significantly less than the decrease
in EE when $(M,K)$ is fixed.
A similar phenomenon is observed when the cell size is fixed, but the PCPs decrease proportionately
(e.g., due to technology scaling).
Numerical results presented in Section \ref{sec_sim} are observed to support the analytical results derived
in other sections.

The important new contributions of this paper are, i) we propose a simple analytical condition to decide if the
system should operate in the massive MIMO or non-massive MIMO regime, ii) we derive tight analytical bounds on the
optimal EE in the non-massive MIMO regime, iii) we derive tight analytical bounds for the optimal EE in the massive MIMO regime under the constraint that the maximum number of users is limited in such a way that the
power consumed for ZF multiuser detection and channel estimation is smaller than the sum power consumed by the BS RF receiver/antennas and the transmitter circuitry at the $K$ UTs, iv) we analyze these bounds to study the impact of changing
system parameters on the optimal EE. We believe that this is the first paper to report such an in-depth
analysis of the impact of system parameters on the optimal EE of uplink multiuser MIMO systems
employing ZF multiuser detection.

\section{System Model}
\label{sys_model}
Consider the uplink of a multi-user massive MIMO system where a
BS having $M$ antennas communicates with $K$ single antenna user terminals (UTs).
Let $x_k$ be the complex information symbol transmitted from the $k$-th user. The signal received at the $m$-th BS antenna is then given by
\begin{eqnarray}
\label{sysmodel}
y_m & = & \sum_{k=1}^K \, h_{k,m} x_k  \, + \, n_m  \,\,\,,\,\,\, m=1,2,\cdots,M
\end{eqnarray}
where $n_m$ is the additive white complex circular symmetric Gaussian noise (AWGN) at the $m$-th receiver,
having zero mean and variance $\sigma^2 = N_0 B$.
Here $B$ is the channel bandwidth (Hz), and $N_0$ W/Hz is the power spectral density of the AWGN.
Here $h_{k,m} = \sqrt{G_c} g_{k,m} \in {\mathbb C}$ denotes the complex channel gain between the
$k$-th UT and the $m$-th BS antenna. Also, $g_{k,m}, k=1,2,\cdots,K, m =1,2,\cdots,M$ are i.i.d. ${\mathcal C}{\mathcal N}(0,1)$ (circular symmetric complex Gaussian
having zero mean and unit variance). Further, $\sqrt{G_c} > 0 $ models the geometric attenuation and shadow fading, and is assumed to be constant over many coherence intervals and known a priori to the BS.\footnote{\footnotesize{We consider a simple model where the attenuation of each user's signal is the same. This is done so as to study the effects of transceiver power consumption on the EE in a standalone manner.
Incorporating different attenuation factors makes it difficult to analyze and draw basic insights.}} The model in (\ref{sysmodel}) is also applicable to wide-band channels where OFDM is used.

Let the average power radiated from each UT be $p_u$ Watt (W).  We consider a channel coherence time of $T_c$ seconds, and therefore the number of channel uses in each coherence interval is $T = B T_c$. A part of the coherence interval is
used for acquiring
channel knowledge at the BS. This is usually done through simultaneous transmission of known pilot/training sequences of length $\tau < T$ from each UT. These sequences are chosen to be orthogonal to each other and also satisfy the average transmit power constraint of $p_u$. Due to the requirement of orthogonality between the pilot sequences we must have $\tau \geq K$. The pilot sequences can be represented by the $K \times \tau$ matrix $\sqrt{\tau p_u} \Phi$ whose $k$-th row is the pilot sequence transmitted from the $k$-th UT. Further, $\Phi  \Phi^H = I_K$. Then, the received pilot matrix is given by
\begin{eqnarray}
\label{Ypeqn}
{\bf Y}_p  & = & \sqrt{\tau p_u} {\bf H} \, {\bf \Phi} \, + \, N_p
\end{eqnarray}
where ${\bf H}$ is the multiuser $M \times K$ channel gain matrix with the channel gain between the $k$-th UT and the $m$-th BS antenna i.e., $h_{k,m}$ as its $(m,k)$-th entry. $N_p$ is the AWGN at the BS receive antennas, with i.i.d. ${\mathcal C}{\mathcal N}(0, N_0 B)$ entries. Since $\Phi$ is known at the BS, using ${\bf Y}_p$ it then finds the
minimum mean squared estimate (MMSE) of the channel gain matrix as
{
\vspace{-3mm}
\begin{eqnarray}
\label{chest_op}
{\widehat {\bf H}} & = & \frac{ \sqrt{\tau p_u}}{N_0 B  \, +  \, \tau p_u} \, {\bf Y}_p  \, \Phi^H.
\end{eqnarray}
}
During the data phase, the BS performs multiuser detection using the Zero-Forcing (ZF) receiver based on the channel estimate ${\widehat {\bf H}}$. Let ${\bf y} = ( y_1, \cdots , y_M)^T$ represent the vector of received symbols at the BS, with $y_m$ being the symbol received at the $m$-th BS antenna. Then, from (\ref{sysmodel}) it follows that
${\bf y}  =   {\bf H} \, {\bf x} \, + \, {\bf n}$
where ${\bf x} \Define (x_1 , \cdots, x_K)^T$ is the vector of symbols transmitted by each UT and ${\bf n} \Define (n_1, \cdots, n_M)^T$ is the vector of noise samples at the BS antennas.
Let ${\widehat x_k}$ be the ZF estimate of the symbol transmitted from the $k$-th UT. Then ${\widehat {\bf x}} \Define ({\widehat x_1} , \cdots, {\widehat x_K})^T$ is given by
{
\vspace{-3mm}
\begin{eqnarray}
\label{ZF_op_eqn}
{\widehat {\bf x}} & = & {\widehat {\bf H}}^{\dagger} \, {\bf y}
\end{eqnarray}
}
where ${\widehat {\bf H}}^{\dagger}  \Define  ({\widehat {\bf H}}^H{\widehat {\bf H}})^{-1}{\widehat {\bf H}}^H$ is the pseudo-inverse of ${\widehat{\bf H}}$ ($(\cdot)^H$ denotes the matrix Hermitian operator).
In this paper, for the ZF detector it is assumed that $M \geq  K+1$.\footnote{\footnotesize{A ZF detector is generally defined for $M \geq K$. Due to the lack of closed form expressions for the exact ergodic sum-rate of ZF receivers in a massive multi-user MIMO channel, we use a lower bound to the sum-rate (as proposed in \cite{HNQ}). However this lower bound on the sum-rate is $0$ when $M = K$. Due to this reason, in this paper we only consider the case where $M > K$. Since, $M = K+1$ offers more degrees of freedom than $M = K$ and therefore a larger array gain, it is expected that for the same sum-rate the required power to be radiated (also the power consumed by the PAs) from the UTs would be less when $M = K + 1$ as compared to when $M = K$. On the other hand due to an extra BS antenna in the $M = K+1$ scenario, the power consumption at the BS is expected to increase slightly. Due to this trade-off we therefore expect that the total EE does not vary much between these two scenarios, i.e., $M=K$ and $M = K+1$.}}
In \cite{HNQ} an achievable spectral efficiency (in bits/s/Hz) for the ZF multiuser detector with MMSE channel estimates is given by
{
{
\vspace{-1mm}
\small
\begin{eqnarray}
\label{ZF_sum_rate}
R & = & K \, {\Big (} 1 -  \frac{\tau}{T}{\Big )} \,  \log_2 {\Big (}1 +  \frac{\tau (M - K) (G_c p_u / N_0 B)^2}{ (K + \tau) (G_c p_u / N_0 B) \, + \, 1}  {\Big )} \,\,,\,\, \mbox{i.e.} \nonumber \\
R & = & K \, {\Big (} 1 -  \frac{\tau}{T}{\Big )} \,  \log_2 {\Big (}1 +  \frac{\tau (M - K) \gamma_u^2}{ (K + \tau) \gamma_u \, + \, 1}  {\Big )}  \nonumber \\
& & \,\,,\,\, \mbox{where} \,\,
\gamma_u \, \Define \, \frac{G_c p_u} {N_0 B}.
\end{eqnarray}
\normalsize}
The energy efficiency depends on the total system power consumption, which includes the power radiated
by the UTs.
We would like to subsequently derive an expression for the energy efficiency as a function of $(M, K, R, \tau)$ and the system parameters. Hence, we need to express the power radiated by each UT as a function of $(M, K, R, \tau)$ and the system parameters.
Towards this end, solving for $\gamma_u$ in (\ref{ZF_sum_rate}) we get\footnote{\footnotesize{{From (\ref{ZF_sum_rate}), we know that the achievable sum rate $R$ is a function of $(\gamma_u, M, K, \tau, T)$.
This then implies that, for a fixed $(M, K, \tau, T)$ and a given desired sum rate $R$, the required power to radiated from each UT must be a function of $(R, M, K, \tau, T)$.
Towards finding this function, we note that from (\ref{ZF_sum_rate}), we get the following quadratic equation in $\gamma_u$, i.e.,
$a_1 \gamma_u^2 - a_2 \gamma_u - a_3  =  0$.
Here $a_1  \, \Define  \, \tau (M - K) \,\,,\,\, a_2 \, \Define \,   (K + \tau) \, {\Big (} 2^{\frac{R}{K(1 - \frac{\tau}{T})} } - 1 {\Big )}  \,\,,\,\,
a_3 \, \Define \,  {\Big (} 2^{\frac{R}{K(1 - \frac{\tau}{T})} } - 1 {\Big )}$.
Out of the two roots of this quadratic equation, one root is positive
and the other is negative. Since $p_u > 0$, $\gamma_u$ must be positive and therefore we only consider the positive root, which is given by
(\ref{ZF_sum_rate_gammau}).}}}
}
{
\small
\begin{eqnarray}
\label{ZF_sum_rate_gammau}
\gamma_u & = & \frac{K + \tau}{2 \tau (M - K)} {\Big (} 2^{\frac{R}{K(1 - \frac{\tau}{T})} } - 1 {\Big )}  \, + \,  \nonumber \\
& &  \hspace{-7mm}  \sqrt{{\Bigg (} \frac{K + \tau}{2 \tau (M - K)} {\Big (} 2^{\frac{R}{K ( 1 - \frac{\tau}{T})} } - 1 {\Big )}{\Bigg )}^2 +  \frac{2^{\frac{R}{K (1 - \frac{\tau}{T})} } - 1}{\tau (M - K)}}
\end{eqnarray}
\normalsize
}
The following lemma gives a useful upper bound on $\gamma_u$.

\begin{mylemma}
\label{lemma_gammau_ub}
For any $(M , K , \tau, R , T)$ such that $M > K$ and $T > \tau \geq K \geq 1$, it follows that{\footnote{\footnotesize{{The proposed upper bound on $\gamma_u$ in Lemma \ref{lemma_gammau_ub} has been used to prove several important results later in the paper (Theorem \ref{thm6_stmt}, Theorem \ref{thm7_stmt}
 and Theorem \ref{thm2_stmt}). 
The bound in Lemma \ref{lemma_gammau_ub} is sufficiently tight in order that these important results hold.
A tighter bound in Lemma \ref{lemma_gammau_ub} (compared to the proposed bound)
will not change the key insights already conveyed by these Theorems.
The results in these Theorems hold under $(R, T)$ and the PCPs satisfying certain conditions.
For a fixed $(R, T)$, further tightening of the bound in Lemma \ref{lemma_gammau_ub} can lead to these results
being valid for a broader range of values of the PCPs. However, for many practical scenarios, this enlargement of the range of values for the PCPs is insignificant.}
}}}
\begin{eqnarray}
\label{eqn_22}
\gamma_u & < & \frac{({ K} \, + \, { \tau}) \,{\Big (}  2^{\frac{R}{{ K} ( 1 - \frac{{ \tau}}{T})}} - 1 {\Big )} }{{ \tau} ({ M} - {K})} \, + \, \frac{1}{{ K} \, + \, { \tau}}.
\end{eqnarray}
\end{mylemma}

{\em Proof}:
Starting with the expression of $\gamma_u$ in (\ref{ZF_sum_rate_gammau}) we have
\begin{eqnarray}
\label{eqn_19_thm}
\gamma_u & = & \frac{({K} + { \tau}) \, {\Big (}  2^{\frac{R}{{ K} ( 1 - \frac{{ \tau}}{T})}} - 1 {\Big )}}{2 { \tau} ({ M} - { K})} \, {\Big (} 1 + \sqrt{1 + v}  {\Big )}  \nonumber \\
v & \Define &  \frac{4 { \tau} ({ M} - { K})}{({ K} + { \tau})^2 \,  {\Big (}  2^{\frac{R}{{ K} ( 1 - \frac{{ \tau}}{T})}} - 1 {\Big )}}.
\end{eqnarray}
Using the fact
$\sqrt{1 + v}  <  1 + \frac{v}{2}$
in (\ref{eqn_19_thm}) we get (\ref{eqn_22}) which finishes the proof.
$\hfill\blacksquare$

\section{Power consumption model and the optimal EE}
\label{sec_pow_model}
In the following we model the power consumed at the UTs and at the BS.
The average power consumed by each user's transmitter can be modeled as $p_{tx} = \alpha p_u \, + \, p_t$ where $\alpha > 1$ models the efficiency of the power amplifier\footnote{\footnotesize{As in other papers \cite{Miao, DHa, HYang, Bjorn}, we also assume that the power amplifiers (PAs) in the UTs operate in the linear region of their transfer characteristic curve, i.e., where doubling the radiated power proportionately doubles the consumed power as well. Numerical results in Section \ref{sec_sim} suggest that by varying $(M,K)$ optimally with changing system parameters, the dynamic range
requirement for the output radiated power level is much smaller when compared to that of systems where $(M,K)$ is fixed.}} and $p_t$ is the power consumed by the other signal processing circuits inside the transmitter (e.g., oscillator, digital-to-analog converter, filters) \cite{Bjorn,cmos_design, Nemati, e_eff_tx}.

At the BS, let $p_r$ (in Watt) be the average power consumed in each BS receiver antenna unit (e.g., per-antenna RF and baseband hardware). The average power consumed at the BS for decoding each user's coded information stream is modeled as $p_{dec}$ (in Watt).\footnote{\footnotesize{With fixed $R$ and varying $K$, it is expected that the per-user information rate ($R/K$) and therefore
$p_{dec}$ will vary. However, in this paper we assume $R$ to be not very large, so that the
variation in $p_{dec}$ is relatively small compared to the value of $p_t$.
Since $p_{dec}$
and $p_t$ impact the PCPs only through their sum,
varying $K$ impacts the PCPs negligibly when $R$ is not very large.}}
{
From Table \ref{table_0} it is clear that the total number of complex operations to be computed (per coherence interval) for channel estimation and ZF multiuser detection
is $(2 M K T \, + \, 4 M K ^2 \, + \, (8 K^3/3))$.
Let $C_0$ Joule (J) denote the energy required to compute a single complex operation.
As these many operations are computed in $T_c$ seconds, the average power consumed for channel estimation and multiuser detection is therefore given by}
\begin{eqnarray}
\label{mud_pow}
p_{mud} & = & 2 M K C_0 B \, + \, 4 M K^2 \frac{C_0}{T_c} \, + \, 8 K^3 \frac{C_0}{3 T_c}.
\end{eqnarray}
\vspace{-5mm}
{
\begin{center}
\small
\begin{table*}
\caption{{Number of complex-valued operations required
for channel estimation and ZF multiuser detection}}
\centering
\footnotesize{
\vspace{-6mm}
{
\begin{tabular}[t]{| c | c | c | c |}
\hline
 & \mbox{Computation} & \mbox{No. of operations} & \mbox{Description} \\ \hline
\hline
A & \mbox{Channel estimation phase ($\tau$ channel uses)} & $2 M K \tau \, + \, 4 M K^2 $  &  A.$1 + $A.$2$ \\ 
& & $\, + \, (8 K^3/3)$ & \\ \hline
A.$1$ & \mbox{Computing the channel estimate ${\widehat {\bf H}}$ } & $2 \, M \, K \, \tau$  & \mbox{Multiplication of a $M \times \tau$ matrix } \\
& & &  with a $\tau \times K$ matrix (see (\ref{chest_op}) and \cite{BoydV}) \\ \hline
A.$2$ & Computing the pseudo-inverse of ${\widehat {\bf H}}$, i.e., & &  \\
&  ${\widehat {\bf H}}^{\dagger}  \Define {\Big (} {\widehat {\bf H}}^H {\widehat {\bf H}} {\Big )}^{-1} \,{\widehat {\bf H}}^H $ & $4 M K^2 \, + \, (8 K^3/3)$ & A.$2$.$1$ + A.$2$.$2$ + A.$2$.$3$ \\
\hline
A.$2$.$1$ & Computing ${\bf A} = {\widehat {\bf H}}^H {\widehat {\bf H}}$ & $2 \, M \, K^2$ & Multiplication of a $K \times M$ matrix with a \\
& & & $M \times K$ matrix \\ \hline
A.$2$.$2$ & Computing ${\bf B} = {\bf A}^{-1}$ & $8 K^3 / 3 $ & Inversion of a $K \times K$ matrix, see \cite{BoydV} \\
\hline
A.$2$.$3$ & Computing ${\widehat {\bf H}}^{\dagger} = {\bf B} {\widehat {\bf H}}^H$ & $2 \, M \, K^2$ & Multiplication of a $K \times K$ matrix with a \\
& & & $K \times M$ matrix \\ \hline
\hline
B & Data phase ($T - \tau$ channel uses) & $2 M K (T - \tau)$  &  Multiplication of a $K \times M$ matrix with a  \\ 
 & ZF multiuser detection &  &  $M \times 1$ vector in every channel use (see (\ref{ZF_op_eqn})) \\  \hline
\hline
C & Channel estimation  + ZF multiuser detection & $2 M K T \, + \, 4 M K^2 $ & A + B \\ 
& & $\, + \, (8 K^3/3)$ & \\ \hline
\end{tabular}
}
\label{table_0}
}
\end{table*}
\vspace{-15mm}
\end{center}
\normalsize
}
Let $p_s$ model the fixed power consumption (e.g. control layer operations, backhaul) which is independent of $M$ and $K$.
Then the total system power consumed (in Watt) is given by
{
\vspace{-2mm}
\begin{eqnarray}
\label{P_eqn0}
P & = & K p_{tx} \, + \, \overbrace{(K p_{dec}  + M p_{r} + p_{mud})}^{\mbox{\small{Power consumed at BS}}}  \, + p_s  \nonumber \\
 & = & K (\alpha p_u + p_t  + p_{dec}) \, + \, M p_r + p_{mud} + p_s.
\end{eqnarray}
}
Note that $p_t$ and $p_{dec}$ contribute to $P$ only through their sum
and therefore for brevity of notation, let
{
\vspace{-2mm}
\begin{eqnarray}
p_d & \Define & p_t \, + \, p_{dec} \,\,\,,\,\,\, \mbox{and therefore} \nonumber \\
P &  =  & K ( \alpha p_u + p_d  ) + M p_{r} + p_{mud} + p_s.
\end{eqnarray}
}
Using the expression for $p_{mud}$ from (\ref{mud_pow}) we further get
{
\vspace{-2mm}
\begin{eqnarray}
\label{P_eqn}
P & = & \alpha K p_u \, + \, p_s  \, + \,  K {\Big (} p_d \, + \, 8 K^2 \frac{C_0}{3 T_c} {\Big )}  \nonumber \\
 & & \,\,\, +  \,  M {\Big (} p_r + 2 K C_0 B + 4 K^2 \frac{C_0}{T_c}{\Big )}.
\end{eqnarray}
}
The EE (bits/Joule) is given by
{
\vspace{-2mm}
\begin{eqnarray}
\label{e_eqn}
\eta_{zf} & = & { R B }/{P}.
\end{eqnarray}
Multiplying (\ref{P_eqn}) by $G_c/(N_0 B)$ on both sides and using the fact that $\gamma_u = G_c p_u /(N_0 B)$ (see (\ref{ZF_sum_rate})) we get
\begin{eqnarray}
\label{p_sum}
\frac{G_c P}{N_0 B} &= & \alpha K \gamma_u \, + \, \rho_s \, + \, K {\Big (} \rho_d \, + \, \frac{8}{3} K^2  \frac{\rho_0}{T} {\Big )}   \nonumber \\
&  & \,\,  +  \,  M {\Big (} \rho_r \, + \, 2 K \rho_0 \, + \, 4 K^2 \frac{\rho_0}{T}{\Big )}
\end{eqnarray}
}
where the normalized PCPs are given by\footnote{\footnotesize{The division of the PCPs $(p_r, p_d, p_s, C_0 B)$ by $N_0 B$ is motivated by the fact that studies have shown that the power consumption in band-limited transceiver circuits is typically proportional to $N_0 B$ (the constant of proportionality depends on technology and design parameters) \cite{Nossek},\cite{UDMCMOS}.}}
{
\begin{eqnarray}
\label{rho_def}
\rho_r & \hspace{-3mm} \Define & \hspace{-3mm} \frac{G_c p_r}{N_0 B} \,,\, \rho_d \Define \frac{G_c p_d}{N_0 B} \,,\, \rho_s \Define \frac{G_c p_s}{N_0 B} \,,\,
\rho_0  \Define  \frac{G_c C_0}{N_0}.
\end{eqnarray}
}
Also, let the normalized EE be given by
{
\vspace{-2mm}
\begin{eqnarray}
\label{zeta_zf_def}
\zeta_{zf} & \Define & \eta_{zf} \frac{N_0}{G_c}
\, = \, \frac{R}{G_c P / (N_0 B)}
\end{eqnarray}
}
where the last step follows from (\ref{e_eqn}).
Using the expression for $G_c P / (N_0 B)$ from (\ref{p_sum}) into (\ref{zeta_zf_def}) we get
\begin{eqnarray}
\label{zeta_zf_1}
\frac{R}{\zeta_{zf}(M, K, \tau, R , \Theta)} & \hspace{-3mm} = & \hspace{-3mm} \alpha K \gamma_u \, + \, \rho_s  +  K {\Big (} \rho_d \, + \, \frac{8}{3} K^2  \frac{\rho_0}{T} {\Big )}   \nonumber \\
 &  &   +  \, M {\Big (} \rho_r \, + \, 2 K \rho_0 \, + \, 4 K^2 \frac{\rho_0}{T}{\Big )}
\end{eqnarray}
where $\Theta \Define {\Big (}\alpha, \rho_r, \rho_d, \rho_s, \rho_0 , T {\Big )}$,
and we use the notation $\zeta_{zf}(M , K, \tau, R , \Theta)$ to explicitly highlight the dependence of $\zeta_{zf}$ on $(M , K, \tau, R , \Theta)$.

In this paper, we are interested in maximizing the EE $\zeta_{zf}(M, K, \tau, R , \Theta)$ as a function of $(M,K,\tau)$ for a given $(R , \Theta)$.
Our aim is to study the impact of $(\rho_r, \rho_d, \rho_s, \rho_0)$
on the optimal EE.
For a given $(R , \Theta)$, the optimal EE $\zeta_{zf}^{\star}(R , \Theta)$ is given by
{
\vspace{-2mm}
\begin{eqnarray}
\frac{1}{\zeta_{zf}^{\star}(R , \Theta)} & = & \min_{\substack{ (M, K, \tau) \in {\mathbb Z}^3 \, \vert \, \\
1 \leq K \leq \tau < T \\
M > K}}   \, \frac{1}{\zeta_{zf}(M , K, \tau, R , \Theta)}.
\end{eqnarray}
}
For a given $(R, \Theta)$ let the optimal $(M,K,\tau)$ be denoted by
{
\vspace{-2mm}
\begin{eqnarray}
\label{opt_int_M_K_tau}
{\Big (} M_{zf}^{\star}(R, \Theta) \,,\, K_{zf}^{\star}(R , \Theta)  \,,\, \tau^{\star}(R , \Theta) {\Big )} &  =  &  \nonumber \\
& & \hspace{-40mm}  \arg \hspace{-3mm} \min_{\substack{ (M, K, \tau) \in {\mathbb Z}^3 \, \vert \, \\
1 \leq K \leq \tau < T \\
M > K}}  \frac{1}{\zeta_{zf}(M, K, \tau, R , \Theta)}.
\end{eqnarray}
}
Note that varying the normalized PCPs can model scenarios where the power consumed by the various
hardware components (e.g., RF receiver at the BS, UT transmitter circuitry, channel decoder at BS, baseband processors) changes due to technology scaling. Since all the normalized PCPs are proportional to the channel gain $G_c$ (see (\ref{rho_def})),
the effect of varying cell size can also be studied.
We firstly show that, irrespective of the fixed value of $(R , T , \alpha)$, the optimal EE decreases with increasing $(\rho_r, \rho_d , \rho_s, \rho_0)$.

\begin{mytheorem}
\label{thm91_stmt}
Consider $\Theta_1 \Define (\alpha, \rho_{r_1}, \rho_{d_1}, \rho_{s_1}, \rho_{0_1}, T)$ and
$\Theta_2 \Define (\alpha, \rho_{r_2}, \rho_{d_2}, \rho_{s_2}, \rho_{0_2}, T)$.
If $\rho_{r_2} \geq \rho_{r_1}$, $\rho_{d_2} \geq \rho_{d_1}$, $\rho_{s_2} \geq \rho_{s_1}$ and $\rho_{0_2} \geq \rho_{0_1}$, with at least one of these being a strict inequality, it follows that
{
\vspace{-3mm}
\begin{eqnarray}
\zeta_{zf}^{\star}(R , \Theta_1) & > & \zeta_{zf}^{\star}(R , \Theta_2).
\end{eqnarray}
}
\end{mytheorem}

{\em Proof}:
Refer to Appendix \ref{prf_thm91_stmt}.
$\hfill\blacksquare$

We divide our analysis of the optimal EE into two parts, depending upon
whether the normalized PCPs are ``large'' or ``small'' for a given $(R , T, \alpha)$.
In this paper, for a given $(R, T, \alpha)$ the PCPs are said to be ``large''
if the optimal $(M^{\star}_{zf}(R, \Theta) \,,\, K^{\star}_{zf}(R, \Theta)) = (2,1)$, and is said to be
``small'' otherwise.\footnote{\footnotesize{{
In Section \ref{subsec_largerho} it is shown that with ``sufficiently large'' values of the normalized PCPs (for a given $(R, T, \alpha)$), it is optimal to have {\em few} BS antennas communicating with a single UT. Therefore the large PCP regime is also referred to as the ``non-massive MIMO'' regime (i.e., few BS antennas and few UTs).
In Section \ref{subsec_smallrho} it is shown that with ``sufficiently small'' values of the normalized PCPs (for a given $(R, T, \alpha)$), it is optimal to have a large number of BS antennas communicating with many UTs. Therefore the small PCP regime is also referred to as the ``massive MIMO'' regime (i.e., a large number of antennas at the BS and many UTs).}}}

\section{Large $(\rho_r, \rho_d, \rho_s, \rho_0)$: Non-massive MIMO regime}
\label{subsec_largerho}

In this section we study those scenarios where the normalized PCPs
take large values (e.g., small cells and/or high power consuming hardware).
The following theorem shows that for these scenarios it is optimal to operate in a single-user mode with few antennas at the BS (i.e., non-massive MIMO regime).
\begin{mytheorem}
\label{thm6_stmt}
If $(R , \Theta)$ satisfy\footnote{\footnotesize{For any real number $x$, $\lfloor x \rfloor$
refers to the greatest integer smaller than or equal to $x$.}}
{
\vspace{-1mm}
\begin{eqnarray}
\label{final_suff_cnd}
(\rho_r \, + \, 2 \rho_0) &  \hspace{-3mm} \geq &  \hspace{-3mm} \frac{\alpha}{1 + \lfloor \sqrt{T} \rfloor }  +  \frac{\alpha (1 + \lfloor \sqrt{T} \rfloor )}{\lfloor \sqrt{T} \rfloor}
{\Big (} 2^{\frac{R}{1 - ({\lfloor \sqrt{T} \rfloor}/{T})}} \, - \, 1 {\Big )}   \nonumber \\
\end{eqnarray}
}
and $T > 1$, then $M_{zf}^{\star}(R, \Theta) \, = \, 2 \, , \, K_{zf}^{\star}(R, \Theta) \, = \, 1$.
\end{mytheorem}

{\em Proof}:
Refer to Appendix \ref{prf_thm6_stmt}.
$\hfill\blacksquare$

\begin{myremark}
\label{remark_large_r}
From Theorem \ref{thm6_stmt} it is clear that for a given $(R,T,\alpha)$ the optimal
$(M_{zf}^{\star}(R, \Theta),K_{zf}^{\star}(R, \Theta) ) = (2,1)$
when $(\rho_r + 2 \rho_0)$ is ``{\em sufficiently}'' large (i.e., greater than the value in the
R.H.S. of (\ref{final_suff_cnd})).
{The inequality in (\ref{final_suff_cnd}) is very important, since it gives us
the insight that it is optimal to operate in the non-massive MIMO regime when
the PCPs are sufficiently large for the given $(R, T, \alpha)$.} 
The normalized PCPs are large either when the cell size is small (i.e., large $G_c$) or when
$(p_r, p_d, p_s, C_0)$ are large (e.g., due to power inefficient RF design). When cell size is small, path loss is less and therefore less power is required to be radiated by the UT, which results in the total power consumption being dominated by the power consumption of sources other than the PA. Due to small path loss, array gain is not really required since the received signal power at the BS antenna is already sufficiently high to support the given information rate. Hence, EE
is maximized by using the least number of BS antennas and the least number of UTs.\footnote{\footnotesize{Based on the discussion above, we believe that the optimal $(M,K)$ for a general ZF precoder
with $M \geq K$ would be $M = K =1$.
However, the importance of Theorem \ref{thm6_stmt} lies
not in showing the exact value of the optimal $(M,K)$, but in the fact that the optimal $(M,K)$
are small compared to their typical values in a massive MIMO scenario.}}

Similarly with increasing $(p_r, p_d, p_s, C_0)$ and fixed channel gain $G_c$, the power consumed by the PA becomes small when compared to the power consumed by the other system components. In such a scenario increasing $M$
to reduce the PA power consumption (i.e., exploiting array gain to reduce the required
radiated power), will result in a much more increase in the power consumed by the BS (due to more RF receivers), as compared to the saving in the PA power consumption. Therefore even for this scenario it is optimal to have $(M,K) = (2,1)$.
These observations have been confirmed through numerical simulations (see Fig.~\ref{rev_fig_1}
in Section \ref{sec_sim}).
$\hfill\blacksquare$
\end{myremark}

The next theorem proposes tight bounds on the optimal EE.
\begin{mytheorem}
\label{thm7_stmt}
Let the unnormalized optimal EE be denoted by
{
\vspace{-1mm}
\begin{eqnarray}
\label{eta_star_def}
\eta_{zf}^{\star}(R , \Theta) & \Define & \frac{G_c}{N_0} \zeta_{zf}^{\star}(R , \Theta).
\end{eqnarray}
}
If $(R, \Theta)$ satisfy (\ref{final_suff_cnd}) and $T > 1$ then
{
\vspace{-1mm}
\begin{eqnarray}
\label{er_def}
\frac{2}{3} \, e(R , \Theta) & < & \eta_{zf}^{\star}(R , \Theta) \, < \, e(R,\Theta) \,\,\,,\,\,\, \mbox{where}  \,\,  \nonumber \\
e(R, \Theta)  &  \Define  &  \frac{R \,  B}{2 p_r + p_d + p_s + 4 C_0 B + \frac{32}{3} \frac{C_0}{T_c}}.
\end{eqnarray}
}
\end{mytheorem}

{\em Proof}:
Refer to Appendix \ref{prf_thm7_stmt}.
$\hfill\blacksquare$

\begin{myremark}
\label{remark_large_r_1}
With large values of the PCPs (either small cell size or inefficient RF design)
it is clear that the total system power consumed will be dominated by the power consumed
in the two BS receivers and the UT transmitter circuitry, i.e., $P \approx (2 p_r + p_d + p_s + p_{mud})$.
Since $(p_r, p_d, p_s, C_0)$ are independent of $G_c$ it follows that with increasing $G_c$, the EE becomes
increasingly insensitive to variations in $G_c$.\footnote{\footnotesize{We would expect a similar result even for the scenario where the channel gains to the users are different. When the cell size is small,
the path loss to all the users is less and therefore none of them need the BS array gain to achieve
their target spectral efficiencies. Since BS array gain is anyways not required, $M$ should
be kept small since it would otherwise increase the BS power consumption.}}
This conclusion is supported by Theorem \ref{thm7_stmt} as both the tight upper and lower bounds
in (\ref{er_def}) are independent of $G_c$ (see also Fig.~\ref{rev_fig_1} in Section \ref{sec_sim}).
Since $P \approx (2 p_r + p_d + p_s + p_{mud})$ for large values of the PCPs, it follows that the EE decreases linearly with increasing $(p_r, p_d, p_s, C_0)$
(see Fig.~\ref{rev_fig_4} in Section \ref{sec_sim}).
$\hfill\blacksquare$
\end{myremark}

{
\begin{myremark}
Both Theorem \ref{thm6_stmt} and \ref{thm7_stmt} are valid when $(R,\Theta)$ satisfy the condition in (\ref{final_suff_cnd}).
We firstly note that the R.H.S. of the condition in (\ref{final_suff_cnd})
is exactly $\alpha $ times the proposed upper bound to $\gamma_u$ in Lemma \ref{lemma_gammau_ub} with $(M=2, K=1, \tau = \lfloor \sqrt{T} \rfloor)$.    
From (\ref{eqn_73}) in Appendix \ref{prf_thm6_stmt} and
step (c) of (\ref{zeta_21_ub}) in Appendix \ref{prf_thm7_stmt} it is clear that
the proof of Theorems \ref{thm6_stmt} and \ref{thm7_stmt} 
uses the proposed upper bound to $\gamma_u$ in Lemma \ref{lemma_gammau_ub} with $(M=2, K=1, \tau = \lfloor \sqrt{T} \rfloor)$.
From the proofs it follows that tightening 
the upper bound in Lemma \ref{lemma_gammau_ub} will lead to a corresponding relaxation of the condition in (\ref{final_suff_cnd}), i.e., the R.H.S. of
the condition in (\ref{final_suff_cnd}) will decrease. 
In the proof of Lemma \ref{lemma_gammau_ub} we have used the bound $\sqrt{1 + v} < 1 + (v/2)$ for any $v > 0$.
This bound becomes increasingly tighter as $v \rightarrow 0$.
For typical values of $(R, T)$ we see that $v$ is very small i.e., the bound in Lemma \ref{lemma_gammau_ub} is tight, and therefore the corresponding relaxation of the condition in
(\ref{final_suff_cnd}) is insignificant.\footnote{\footnotesize{{From (\ref{eqn_19_thm}) in the proof of Lemma \ref{lemma_gammau_ub} it follows that for $(M=2, K=1, \tau = \lfloor \sqrt{T} \rfloor)$, we have $v =  4 \lfloor \sqrt{T} \rfloor / {\Big (} (1 + \lfloor \sqrt{T} \rfloor)^2 \, {\Big (} 2^{\frac{R}{( 1 - \frac{{\lfloor \sqrt{T} \rfloor }}{T})}} - 1  {\Big )} {\Big )}$.
With typical values of $R=8$ bps/Hz, $B = 200$ KHz, $T_c = 2$ ms, i.e., $T = B T_c = 400$, we have $v = 5.3 \times 10^{-4}$.
The difference between the upper bound and the exact value of $\sqrt{1 + v}$ (i.e.,
$1 + v/2 - \sqrt{1 + v}$) is roughly of the order of $10^{-7}$.
This implies that with $(R=8, T=400)$, the condition in (\ref{final_suff_cnd})
can be relaxed by reducing the R.H.S. of (\ref{final_suff_cnd}) at most by a factor of $(1 + 10^{-7})$.
}}}
$\hfill\blacksquare$
\end{myremark}
}

\section{Small $(\rho_r, \rho_d, \rho_s, \rho_0)$: Massive MIMO regime}
\label{subsec_smallrho}
In this section, we study the impact of the normalized PCPs
on the optimal EE when these parameters take small values (e.g., large cell size and/or
low power consuming hardware).
Numerical simulations reveal that the optimal $(M,K, \tau)$ are large when the normalized PCPs take small values.
This is why we refer to this scenario as the massive MIMO regime.

In reality, physical channels have finite dimensionality \cite{Burr, Muller} (for e.g., due to insufficient scattering).
With finite dimensional channels it is not possible to spatially multiplex a large number of users, and therefore
in this paper we consider scenarios where the maximum number of supported users is in a few tens.
With a few tens of users it turns out that with current technology the power consumed for
channel estimation and multiuser detection is smaller than the sum power consumed by the $M$ RF
antenna receivers at the BS, the transmitter circuitry in the $K$ UTs and the $K$ channel decoders in the BS, i.e., $p_{mud} < (M p_r + K p_d)$.
As an example, let us consider a massive MIMO system with $K=20$ users and $M > 20$.
With $B = 200$ KHz, $C_0 = 1 \times 10^{-9}$ Joule, $p_r = p_d = 0.01$ W, $T_c = 2$ ms, we have
$(M p_r + K p_d) = (20 + M) \times 10^{-2}$ W, and $p_{mud} =  (1.07 + 0.88 M) \times 10^{-2}$ W, i.e., $(M p_r + K p_d) \, > \, p_{mud}$ {\em irrespective} of the value of $M$.\footnote{\footnotesize{With $B = 200$ KHz, typical values for $(p_r, p_d, p_s)$ are
in the range $0.01 - 1.0$ W \cite{e_eff_tx}, and that for $C_0$ are less than a nano Joule \cite{green_list}.}}

In Section \ref{sec_theta_cond}, for a given $\Theta$ we will propose a constraint on the maximum possible number of users,
in order that $p_{mud} \leq (M p_r + K p_d)$. Conditions will also be proposed for the PCPs in order that
the maximum possible number of allowed users is larger than ten. Through numerical examples it will be shown that
with current technology, these conditions are usually met.
Thereafter, in Section \ref{sec_analyze_smallrho} we will analyze the optimal EE under this realistic constraint on the
maximum possible number of users.

\subsection{Maximum number of users $K_{max}(\Theta)$, such that $p_{mud} \leq (M p_r + K p_d)$ for all $1 \leq K \leq K_{max}(\Theta)$ and any $M$}
\label{sec_theta_cond}
\begin{mylemma}
\label{lemma_theta_cnds}
Let
{
\vspace{-2mm}
\begin{eqnarray}
\label{kmax_def}
K_{max}(\Theta) & \Define & \min{\Big (}  \frac{T}{4} \,,\, \frac{\rho_r}{3 \rho_0} \,,\, \frac{3 \rho_d}{2 \rho_0} \, {\Big )}.
\end{eqnarray}
}
Then for any $\Theta$ satisfying the following conditions
{
\begin{eqnarray}
\label{theta_cnds}
\min{\Big (}  \frac{T}{4} \,,\, \frac{\rho_r}{3 \rho_0} \,,\, \frac{3 \rho_d}{2 \rho_0} \, {\Big )} & > & 10 \,\,\,,\,\,\, \mbox{(C.1)} \nonumber \\
\frac{\rho_s}{\alpha} & > & \frac{1}{2} \,\,\,\,,\,\,\, \mbox{(C.2)}
\end{eqnarray}
}
it follows that $K_{max}(\Theta) > 10$.
Further, for all $1 \leq K \leq K_{max}(\Theta)$ and any $M$ we have
\begin{eqnarray}
\label{pmud_dom}
p_{mud} & \leq & (M p_r + K p_d) \,\,,\,\, \mbox{i.e.,}  \nonumber \\
M {\Big (} 2 K \rho_0 \, + \, 4 K^2 \frac{\rho_0}{T} {\Big )} \, + \, K^3 \frac{8 \rho_0}{3T} &  \leq  & (M \rho_r + K \rho_d).
\end{eqnarray}
\end{mylemma}

{\em Proof}:
Refer to Appendix \ref{prf_lemma_theta_cnds}.
$\hfill\blacksquare$

\begin{myremark}
In Lemma \ref{lemma_theta_cnds} it is interesting to note that only a upper limit on $K$ is sufficient
and that no condition is needed on $M$.
The condition (C.1)
depends on the ratio between the PCPs, which does not depend on $G_c$.
Similarly, any change in technology (e.g., device scaling) will have an almost similar impact on all the PCPs, and therefore the condition in (C.1) is expected to still hold.
The satisfiability of condition (C.1) is therefore robust to changing $G_c$ and technology.

The condition (C.2) in (\ref{theta_cnds}) is not required to show (\ref{pmud_dom}).
However we keep (C.2) since it is valid in practical scenarios and also because it will
be useful later in deriving tight bounds on the optimal EE.
$\hfill\blacksquare$
\end{myremark}

With current technology, the conditions in (\ref{theta_cnds}) are generally
satisfied, and the value of $K_{max}(\Theta)$ is in a few tens.
As an example, with $B = 200$ KHz, $C_0 = 1 \times 10^{-9}$ Joule, $p_r = p_d = p_s = 0.1$ W, $T_c = 2$ ms, we have
$K_{max}(\Theta) = 100$ and (C.1) is satisfied.
The condition (C.2) is also satisfied.\footnote{\footnotesize{Power amplifiers used in UTs generally have a power efficiency greater than $5$ percent, i.e., $\alpha < 20$ in most practical scenarios. The fixed power consumption $p_s$ is generally of the order of $1$ W. Even if we consider a worst case scenario with $p_s = 0.1$ W and $G_c = 10^{-12}$ ($120$ dB path loss), we get $\rho_s/\alpha = 6.28 > 1/2$.}}

\subsection{Analysis of (\ref{opt_int_M_K_tau}) for $\Theta$ satisfying (\ref{theta_cnds}) and $1 \leq K \leq K_{max}(\Theta)$}
\label{sec_analyze_smallrho}

From (\ref{opt_int_M_K_tau}) it follows that the optimal EE under the additional constraint $K \leq K_{max}(\Theta)$
is given by
\begin{eqnarray}
\label{prob_primeprime}
& &  \hspace{-65mm} \zeta_{zf}^{\prime\prime}(R , \Theta)  \,  \Define  \,  \zeta_{zf}(M_{zf}^{\prime\prime}(R , \Theta) , K_{zf}^{\prime\prime}(R , \Theta) , \tau^{\prime\prime}(R , \Theta) , R, \Theta)  \nonumber \\
{\Big (} M_{zf}^{\prime\prime}(R , \Theta) , K_{zf}^{\prime\prime}(R , \Theta) , \tau^{\prime\prime}(R , \Theta)  {\Big )} & = &   \nonumber \\
& &  \hspace{-42mm} \arg \hspace{-4mm} \min_{\substack{ (M, K, \tau) \in {\mathbb Z}^3 \, \vert \, \\
1 \leq K \leq K_{max}(\Theta) \,,\, \\
K \leq \tau < T \,,\,
M > K}}  \frac{1 }{\zeta_{zf}(M , K, \tau, R , \Theta)}.
\end{eqnarray}
Unlike section \ref{subsec_largerho}, for small values of the normalized PCPs it appears difficult to solve
the optimization problem in (\ref{prob_primeprime}) exactly.
From numerical simulations it is observed that for small values of the normalized PCPs, the optimal $(M,K,\tau)$ are large (this observation is like a converse to Theorem \ref{thm6_stmt}).  
With large $(M,K,\tau)$, from the expression for $\zeta_{zf}(M, K, \tau, R , \Theta)$ in (\ref{zeta_zf_1}) it is expected that the
relative difference in the EE due to increasing/decreasing $(M,K,\tau)$ by one, i.e., $\vert \zeta_{zf}(M, K, \tau, R , \Theta) - \zeta_{zf}(M\pm 1, K\pm 1, \tau \pm 1, R , \Theta)  \vert / \zeta_{zf}(M, K, \tau, R , \Theta)$ is small. Therefore for small $(\rho_r, \rho_d, \rho_s, \rho_0)$ a good approximation to the optimal EE can be obtained by relaxing the integer constraint on $(M, K, \tau)$ in (\ref{prob_primeprime}) (in Fig.~\ref{rev_fig_1} of Section \ref{sec_sim}, the curves marked with `$<$' and `O' are close together when $(M,K)$ are large).
Let us denote the EE obtained with this relaxation by $\zeta_{zf}^{\prime}(R , \Theta)$, where
\begin{eqnarray}
\label{prob_prime}
 & & \hspace{-67mm} \zeta_{zf}^{\prime}(R , \Theta)  \, \Define \,  \zeta_{zf}(M_{zf}^{\prime}(R , \Theta) , K_{zf}^{\prime}(R , \Theta) , \tau^{\prime}(R , \Theta) , R, \Theta)  \nonumber \\
{\Big (} M_{zf}^{\prime}(R , \Theta) , K_{zf}^{\prime}(R , \Theta) , \tau^{\prime}(R , \Theta)  {\Big )}   &  =   &  \nonumber \\
& &  \hspace{-43mm} \arg \hspace{-4mm} \min_{\substack{ (M, K, \tau) \in {\mathbb R}^3 \, \vert \, \\
1 \leq K \leq K_{max}(\Theta) \,,\, \\
K \leq \tau < T \,,\,
M > K}}  \frac{1 }{\zeta_{zf}(M , K, \tau, R , \Theta)}.
\end{eqnarray}
From Lemma \ref{lemma_theta_cnds} we know that under the conditions in (\ref{theta_cnds}) $p_{mud} \leq (M p_r + K p_d)$
for all $1 \leq K \leq K_{max}(\Theta)$. Further, from the conditions in (\ref{theta_cnds}) we have $K \leq K_{max}(\Theta) \leq T/4$, which implies the availability of abundant channel resource for acquiring accurate channel estimates.
We would therefore expect that with $\Theta$ satisfying the conditions in (\ref{theta_cnds}) and $1 \leq K \leq K_{max}(\Theta)$, $\zeta_{zf}^{\prime}(R , \Theta)$ would be close to the optimal EE of an ideal system
where we assume $p_{mud} = 0$ and perfect channel estimates.
For a given $(R, \Theta)$ the EE of such an ideal system depends only on $(M , K, R, \Theta)$ and is given by
\begin{eqnarray}
\label{eqn_332}
\frac{1}{\zeta_{csi}(M , K, R, \Theta)} & = &  \frac{1}{R} \, {\Bigg [}  \frac{ \alpha K }{(M - K)} {\Big (} 2^{\frac{R}{K}}  - 1  {\Big )}  \nonumber \\
& &  \,\,\,\,\,\,\,\, + \, M \rho_r + K \rho_d + \rho_s {\Bigg ]}.
\end{eqnarray}
Here we have used the fact that for a fixed $(R, M , K)$ with perfect CSI
(i.e., $\tau \rightarrow \infty$, $T \rightarrow \infty$ with $K \leq \tau < T$)
, it is possible to increase $\tau$ in such a manner that we have \cite{HNQ}
\begin{eqnarray}
\label{lim_gammau}
\lim_{\substack{T \rightarrow \infty, \tau \rightarrow \infty , \\
K \leq \tau < T}} \, \gamma_u & = & \frac{(2^{R/K} - 1)}{(M - K)}.
\end{eqnarray}
While evaluating the limit above, we have used the R.H.S. of (\ref{ZF_sum_rate_gammau}) as the expression for $\gamma_u$
and taken $\tau = \sqrt{KT}$.
The optimal EE of such an ideal system is given by
\begin{eqnarray}
\label{zf_csi_opt_MK}
{\Big (} M^{\prime}_{csi}(R , \Theta) \,,\, K^{\prime}_{csi}(R , \Theta) {\Big )} & &  \nonumber \\
&  \hspace{-57mm} \Define & \hspace{-23mm}  \arg \min_{\substack{(M,K) \in {\mathbb R}^2 \, \vert \, \\
1 \leq K \leq K_{max}(\Theta) \, \\
M > K}}  \frac{1}{\zeta_{csi}(M , K, R, \Theta)} \nonumber \\
& & \hspace{-50mm} \zeta_{csi}^{\prime}(R , \Theta)  \,  \Define  \, \zeta_{csi}(M^{\prime}_{csi}(R , \Theta) \,,\, K^{\prime}_{csi}(R , \Theta) , R , \Theta).
\end{eqnarray}
In the following, in Theorem \ref{thm2_stmt} we show that for small values of $(\rho_r, \rho_d, \rho_s, \rho_0)$, $\frac{3}{8} < \frac{\zeta_{zf}^{\prime}(R, \Theta)}{\zeta_{csi}^{\prime}(R, \Theta)} < 1$.
Through numerical simulations it has been observed that the lower bound is tighter than $3/8$.
This is even true when $(M,K,\tau)$ are restricted to be integers (see Fig.~\ref{rev_fig_1}).
This bound is important because (\ref{zf_csi_opt_MK}) can be solved exactly (see Theorem \ref{thm1_stmt} in Appendix \ref{appendix_prev_work}) and analytical expressions
can be derived for $\zeta_{csi}^{\prime}(R , \Theta)$.
We later study
the variation of $\zeta_{csi}^{\prime}(R , \Theta)$ and the optimal $(M,K)$ for the ideal system w.r.t.
varying $(\rho_r, \rho_d, \rho_s, \rho_0)$ but fixed $(R, T, \alpha)$.
Since $\zeta_{csi}^{\prime}(R, \Theta)$ tightly bounds $\zeta_{zf}^{\prime}(R, \Theta)$ we expect that
the variation in $\zeta_{zf}^{\prime}(R, \Theta)$ (and therefore $\zeta_{zf}^{\prime\prime}(R , \Theta)$) with varying $(\rho_r, \rho_d, \rho_s, \rho_0)$ is similar to that
of $\zeta_{csi}^{\prime}(R, \Theta)$. This is verified through exhaustive simulations.

In this section,
we specifically consider situations where the normalized PCPs decrease in
such a way that the ratio between them remains constant, i.e., they still satisfy the conditions in (\ref{theta_cnds}). This models situations where $G_c$ is fixed, but due to technology scaling
all the normalized PCPs decrease by the same factor. This makes sense
since all the sources of power consumption are inherently dependent on the same underlying
fabrication technology. Since all the normalized PCPs are proportional to $G_c$ (see (\ref{rho_def})),
fixed ratio between the PCPs could also model situations where the fabrication technology remains same but the cell size increases (which
reduces $G_c$).
The fixed ratios between the normalized PCPs is denoted by
\begin{eqnarray}
\label{fixed_ratio}
\beta & \Define & \frac{\rho_d}{\rho_r} = \frac{p_d}{p_r} \,,\, \delta \, \Define \, \frac{\rho_s}{\rho_r} = \frac{p_s}{p_r} \,,\,
\mu  \Define  \frac{\rho_0}{\rho_r} = \frac{C_0 B}{p_r}.
\end{eqnarray}
In this section we consider a fixed $(\alpha, R , T , \beta, \delta, \mu)$
satisfying
{
\vspace{-2mm}
\begin{eqnarray}
\label{theta_cnds_1}
\min{\Bigg (} \frac{T}{4} \,,\, \frac{1}{3 \mu} \,,\, \frac{3 \beta}{2 \mu} {\Bigg )} & > & 10
\end{eqnarray}
}
and $\rho_r$ satisfying
{
\vspace{-1mm}
\begin{eqnarray}
\label{theta_cnds_2}
\rho_r  & > &  \frac{\alpha}{2 \delta}   \hspace{46mm} \mbox{(D.1)} \nonumber \\
\rho_r  & > &  {\Bigg (} \frac{3 \alpha}{4 (1 + \beta)^2 \, R} g^2{\Big (} \frac{4 R}{3 K_{max}(\Theta)}{\Big )} {\Bigg )} \,\,\, \mbox{(D.2)} \nonumber \\
\rho_r & < &  \frac{\alpha}{(1 + \beta)^2 } \frac{g^2(R)}{R} \hspace{27mm} \mbox{(D.3)}
\end{eqnarray}
}
where
{
\vspace{-2mm}
\begin{eqnarray}
\label{gx_def_eqn}
g(x) & \hspace{-3mm} \Define & \hspace{-3mm} \sqrt{\frac{x}{2^x - 1}} {\Big (}  2^x \, x \, \log(2) \, - \, 2^x \, + \, 1 {\Big )} \,\,\,,\,\,\, x \geq 0.
\end{eqnarray}
}
The condition in (\ref{theta_cnds_2}) is valid in many practical scenarios.
As an example, with $R = 8$ bps/Hz, $p_r = p_d = 0.01$ W, $p_s = 0.1$ W, $B = 200$ KHz, $T_c = 2$ ms, $\alpha = 2$ and $C_0 = 10^{-9}$ J,
we have $K_{max}(\Theta) = 16.6$
and the condition in (\ref{theta_cnds_2}) is satisfied if and only if $ 0.1 < \rho_r < 2.66 \times 10^{3}$
which corresponds to $G_c$ lying between
$-97$ dB and $-141$ dB.
Note that since $K_{max}(\Theta) > 10$ and $g^2(x)/x$ is strictly monotonically increasing with $x > 0$,
it follows that the R.H.S. of (D.3) is greater than the R.H.S. of (D.2).
{The importance of the conditions in (\ref{theta_cnds_1}) and
(\ref{theta_cnds_2}) stems from the fact that under these conditions our analysis suggests
that it is optimal to operate in the massive MIMO regime (see Theorem \ref{decrhor_incMK}, and Lemma \ref{lemma_918} in Appendix \ref{appendix_prev_work}).}

The following lemma shows that (D.2) in (\ref{theta_cnds_2}) is equivalent to $R < 3R_{max}(\Theta)/4$ where $R_{max}(\Theta)$
depends only on $\Theta$ and is defined in the following lemma.
\begin{mylemma}
\label{lemma_37}
Any $\Theta$ satisfies (\ref{theta_cnds}) if and only if it satisfies
both (\ref{theta_cnds_1}) and (D.1) of (\ref{theta_cnds_2}).
Further any $(R, \Theta)$ satisfies (D.2) of (\ref{theta_cnds_2}) if and only if
$R < 3R_{max}(\Theta)/4$,
where
{
\vspace{-2mm}
\begin{eqnarray}
\label{suff_cnd}
R_{max}(\Theta) & \Define & c(\Theta) \, K_{max}(\Theta) \,\,\, \mbox{where $c(\Theta)$ uniquely satisfies}  \nonumber \\
\frac{g(c(\Theta))}{\sqrt{c(\Theta)}} & = & {\Big (} 1 + \beta {\Big )} \, \sqrt{\frac{K_{max}(\Theta) \rho_r}{\alpha}}.
\end{eqnarray}
}
\end{mylemma}

{\em Proof}:
Refer to Appendix \ref{prf_lemma_37}.
$\hfill\blacksquare$

The following theorem derives tight bounds on $\zeta_{zf}^{\prime}{ (} R , \Theta {)}$ in terms of $\zeta_{csi}^{\prime}{ (} R , \Theta { )}$.
\begin{mytheorem}
\label{thm2_stmt}
For any given $(R , \Theta)$ satisfying both the conditions in (\ref{theta_cnds_1}) and (\ref{theta_cnds_2}) we have
{
\vspace{-2mm}
\begin{eqnarray}
\label{lemma_thm2_eqn}
\frac{3}{8} & < \,  \frac{\zeta_{zf}^{\prime}{\Big (} R , \Theta {\Big )} } {\zeta_{csi}^{\prime}{\Big (} R , \Theta {\Big )}} \,  < &  1.
\end{eqnarray}
}
\end{mytheorem}

{\em Proof:}
The bounds in (\ref{lemma_thm2_eqn}) follow directly from Lemmas \ref{lemma_lb} and \ref{lemma_ub} in Appendix \ref{appendix_prev_work}.{\footnote{\footnotesize{{Lemma \ref{lemma_lb} holds for any $(R, \Theta)$ satisfying
the conditions in (\ref{theta_cnds_1}) and (\ref{theta_cnds_2}).
The proof of Lemma \ref{lemma_lb} uses the proposed upper bound on $\gamma_u$ in Lemma \ref{lemma_gammau_ub}.
Tightening of the upper bound in Lemma \ref{lemma_gammau_ub}
can help in relaxing condition (D.1) in (\ref{theta_cnds_2}) (since $\delta = \rho_s/\rho_r$, this condition is equivalent to $\rho_s > \alpha/2$).
To be precise, for a given $\alpha$ the condition $\rho_s > \alpha/2$
can be relaxed to $\rho_s > \epsilon \alpha/2$ for some $0 < \epsilon < 1$.
However, for many practical scenarios, $\rho_s$ is anyways greater than $\alpha/2$ (see
footnote $15$), and therefore the relaxation $\rho_s > \epsilon \alpha/2$ is not of much significance in these scenarios. More detailed discussion can be found in the proof of Lemma \ref{lemma_lb} in Appendix \ref{appendix_prev_work}.
}}}
}
$\hfill\blacksquare$
 
The following corollary to Theorem \ref{thm2_stmt} proposes bounds on $\zeta_{zf}^{\prime}{\Big (} R , \Theta {\Big )} $ for any $R > 0$ with $\Theta$ required to satisfy (\ref{theta_cnds_1}) and
only condition (D.1) of (\ref{theta_cnds_2}).

\begin{corollary}
\label{corr_9292}
[Corollary to Theorem \ref{thm2_stmt}]

For $R > 0$ and $\Theta$ satisfying (\ref{theta_cnds_1}) and (D.1) of (\ref{theta_cnds_2}), we have
{
\vspace{-2mm}
\begin{eqnarray}
\label{eqn_175}
\frac{3}{8} \, \zeta_{csi}^{\prime}{\Big (} \frac{4}{3}R , \Theta {\Big )} & < \, \zeta_{zf}^{\prime}{\Big (} R , \Theta {\Big )} \, < & \zeta_{csi}^{\prime}{\Big (} R , \Theta {\Big )}
\end{eqnarray}
}
\end{corollary}

{\em Proof}:
The upper bound in (\ref{eqn_175}) follows from the upper bound in Lemma \ref{lemma_ub},
which only requires $\Theta$ to satisfy (\ref{theta_cnds_1}).
The lower bound in (\ref{eqn_175}) follows from step (d) of (\ref{lb_seq}) in Lemma \ref{lemma_lb}.
Note that steps (a) through (d) of (\ref{lb_seq}) require $\Theta$ to only satisfy (\ref{theta_cnds_1})
and (D.1) of (\ref{theta_cnds_2}).
$\hfill\blacksquare$

It can also be shown that for any $(R, \Theta)$ satisfying (\ref{theta_cnds_1})
and (\ref{theta_cnds_2}), a near-optimal solution to the optimal EE $\zeta_{zf}^{\prime}(R , \Theta)$ is obtained
by choosing $M = M_{csi}^{\prime}(4R/3 , \Theta) , K = K_{csi}^{\prime}(4R/3, \Theta), \tau = K_{max}(\Theta)$, i.e.
{
\vspace{-2mm}
\begin{eqnarray}
\zeta_{zf}^{\prime}(R , \Theta) &  > &   \zeta_{zf}{\Big (} M_{csi}^{\prime}(4R/3 , \Theta) \,,\, K_{csi}^{\prime}(4R/3, \Theta)  \nonumber \\
& &  \hspace{20mm} \,,\, K_{max}(\Theta) \,,\, R \,,\, \Theta {\Big )}  \nonumber \\
&  >  &  \frac{3}{8} \zeta_{csi}^{\prime}(4R/3 , \Theta).
\end{eqnarray}
}
Theorem \ref{thm2_stmt} implies that a lot of insights about the variation in $\zeta_{zf}^{\prime}(R , \Theta)$ with
changing $(p_r, p_d, p_s, C_0, G_c)$, can be inferred by studying the corresponding variation in $\zeta_{csi}^{\prime}(R , \Theta)$.
Therefore, in the following we study the impact of varying $(p_r, p_d, p_s, C_0, G_c)$ on $\zeta_{csi}^{\prime}(R , \Theta)$.

The following Corollary to Theorem \ref{thm91_stmt} shows that
when $(p_r, p_d, p_s, C_0)$ scale proportionately (e.g., technology scaling), then
for a fixed $(G_c, \alpha, R , T)$ the unnormalized optimal EE $\eta_{zf}^{\star}(R, \Theta)$
increases {\em strictly} with decreasing $(p_r, p_d, p_s, C_0)$ irrespective of whether we are
in the massive MIMO or the non-massive MIMO regime. A similar result can be shown to hold also for $\eta_{zf}^{\prime}(R, \Theta) \Define G_c \zeta_{zf}^{\prime}(R, \Theta) / N_0$.

\begin{corollary}
\label{corr_9191}
[Corollary to Theorem \ref{thm91_stmt}]\footnote{\footnotesize{{It is to be noted that
proportional scaling of the unnormalized PCPs $(p_r, p_d, p_s, C_0 B)$ with fixed $(R, \alpha, G_c, T)$ is a special case of independent scaling of the normalized parameters $(\rho_r, \rho_d, \rho_s, \rho_0)$
considered in Theorem \ref{thm91_stmt}. Since we discuss this important special case in detail in Section \ref{subsec_smallrho}, Corollary \ref{corr_9191}
has not been placed immediately after Theorem \ref{thm91_stmt} in Section \ref{sec_pow_model}.}}}

Consider a constant $(\alpha, R , T, G_c, \beta, \delta, \mu)$
and a varying $p_r$.
Let $\Theta_1 \Define (\alpha, \rho_{r_1}, \rho_{d_1}, \rho_{s_1}, \rho_{0_1}, T)$ and
$\Theta_2 \Define (\alpha, \rho_{r_2}, \rho_{d_2}, \rho_{s_2}, \rho_{0_2}, T)$.
If $\rho_{r_2} > \rho_{r_1}$ then
{
\vspace{-1mm}
\begin{eqnarray}
\label{eqn_195}
\eta_{zf}^{\star}(R, \Theta_1) & > & \eta_{zf}^{\star}(R, \Theta_2)
\end{eqnarray}
}
where $\eta_{zf}^{\star}(R, \Theta)$ is given by (\ref{eta_star_def}).
\end{corollary}

{\em Proof}:
With $\rho_{r_2}$ strictly greater than $\rho_{r_1}$ and constant $(\beta, \delta, \mu)$, it follows that $\rho_{d_2} > \rho_{d_1}$, $\rho_{s_2} > \rho_{s_1}$ and $\rho_{0_2} > \rho_{0_1}$, and
therefore the condition
for Theorem \ref{thm91_stmt} is satisfied.
Hence from Theorem \ref{thm91_stmt} we have $\zeta_{zf}^{\star}(R, \Theta_1)  >  \zeta_{zf}^{\star}(R, \Theta_2)$, which then implies (\ref{eqn_195}) since $G_c$ is constant.
$\hfill\blacksquare$

The following lemma will be useful later.

\begin{mylemma}
\label{lemma_654}
With decreasing $\rho_r$ and fixed $(\alpha, R , T , \beta, \delta, \mu)$,
$x^{\prime}(R, \Theta)$ decreases strictly monotonically.
\end{mylemma}

{\em Proof}:
Follows from the expression of $g(x^{\prime}(R , \Theta))$ in (\ref{gxprime_eqn_rev}),
and the facts that, i) $g(x)$ is strictly monotonically increasing with $x > 0$, and
ii) $\rho_d/\rho_r = \beta$ is constant.
$\hfill\blacksquare$

The following Theorem shows that for a fixed $(\alpha, R , T, p_r, p_d, p_s, C_0)$ and
varying $G_c$, the unnormalized optimal EE of the ideal system i.e., $\eta_{csi}^{\prime}(R, \Theta)$
decreases {\em strictly} monotonically with decreasing channel gain $G_c$ when $G_c$ is sufficiently small.
Numerical simulations reveal a similar behaviour for $\eta_{zf}^{\prime}(R , \Theta) = G_c \zeta_{zf}^{\prime}(R , \Theta)/N_0$ (see Fig.~\ref{rev_fig_1} in Section \ref{sec_sim}).

\begin{mytheorem}
\label{decGc_deceta}
Consider a constant $(\alpha, R , T, p_r, p_d, p_s, C_0)$ satisfying
(\ref{theta_cnds_1})
and $G_c$ satisfying
{
\vspace{-2mm}
\begin{eqnarray}
\label{cnd_23}
G_c  & > & \frac{N_0 B }{p_r} \, \frac{\alpha}{2 \delta}  \,\,,\,\, \nonumber \\
G_c & > &  \frac{N_0 B }{p_r} \, {\Bigg (} \frac{3 \alpha}{4 (1 + \beta)^2 \, R} g^2{\Big (} \frac{4 R}{3 K_{max}(\Theta)}{\Big )} {\Bigg )} \,\,,\,\,  \nonumber \\
G_c & < & \frac{N_0 B }{p_r} \, \frac{\alpha}{(1 + \beta)^2 \, R} g^2(R).
\end{eqnarray}
}
Let $\eta_{csi}^{\prime}(R , \Theta) \Define G_c \zeta_{csi}^{\prime}(R , \Theta) / N_0$
be the unnormalized optimal EE of the ideal system. Then $\frac{\partial \eta_{csi}^{\prime}(R , \Theta)}{\partial G_c} \, > \, 0$.
\end{mytheorem}
{\em Proof}:
Refer to Appendix \ref{prf_decGc_deceta}.
$\hfill\blacksquare$

The following theorem shows that with constant $(\alpha, R, T, \beta, \delta, \mu)$
and $(R, \Theta)$
satisfying (\ref{theta_cnds_1}) and (\ref{theta_cnds_2}), the optimal $(M,K)$ for the
ideal system increases monotonically with decreasing $\rho_r$.

\begin{mytheorem}
\label{decrhor_incMK}
For a constant $(\alpha, R, T, \beta, \delta, \mu)$ and $(R, \Theta)$
satisfying the conditions in (\ref{theta_cnds_1}) and (\ref{theta_cnds_2})
it follows that both $K^{\prime}_{csi}(R , \Theta)$ and $M^{\prime}_{csi}(R, \Theta)$ increase monotonically
with decreasing $\rho_r$.
\end{mytheorem}

{\em Proof}:
Refer to Appendix \ref{prf_decrhor_incMK}.
$\hfill\blacksquare$

\begin{myremark}
\label{remark_small_r}
The results of Corollary \ref{corr_9191}, Theorem \ref{decGc_deceta} and Theorem \ref{decrhor_incMK}
is discussed in the following.
In all these results $(\alpha, R, T)$ is fixed.
We firstly consider the scenario where $(p_r,p_d, p_s, C_0)$ is fixed and $G_c$ is decreasing (i.e., increasing
cell size). Starting with a sufficiently large $G_c$, we know from Theorem \ref{thm6_stmt} that the optimal
$(M, K) = (2,1)$ (i.e., non-massive MIMO regime). With decreasing $G_c$ it is expected that the UT would be required to increase its radiated power
linearly so as to achieve a constant $R$. This increase will increase the power consumed by the PA
in the UT until the power consumed by the PA dominates the total system power consumption.
Therefore, with further decrease in $G_c$ the EE will start decreasing linearly with $G_c$, i.e. a $20$ dB
increase in path loss will reduce the EE by a factor of roughly $100$.

To reduce the amount of loss in EE, the system must increase $(M,K)$ so as to reduce
the required power to be radiated by the UTs, by exploiting array and multiplexing gains.\footnote{\footnotesize{With near perfect CSI (since $K \leq K_{max}(\Theta) \leq T/4$), the total power
consumed by the $K$ PAs is $\alpha K p_u \approx \alpha K \frac{(2^{R/K} -1)}{(M - K)} \frac{N_0 B}{G_c}$ (see (\ref{lim_gammau})), where $(M - K)$ in the denominator models the array gain due to the extra $(M - K)$ degrees of freedom. For a fixed $(M,K)$, it is clear that the power consumed by the PAs increases as $1/G_c$ with decreasing $G_c$. However if we increase both $M$ and $K$ with decreasing $G_c$ in such a way that $(M - K)$ increases, then it is clear that
the total power consumed by the $K$ PAs would increase at a rate slower than $1/G_c$, since $K (2^{R/K} - 1)$ decreases monotonically with increasing $K$ (fixed $R$).
}} However $(M,K)$
must be increased in a controlled manner since the power consumed at the BS would also increase with increasing
$(M,K)$. Numerical simulations suggest that increasing $(M,K)$ indeed
reduces the amount of loss in EE when compared to fixed $(M,K) = (2,1)$ scenario (see Fig.~\ref{rev_fig_1} of Section \ref{sec_sim}). This increase in $(M,K)$ with decreasing $G_c$ (and therefore decreasing $(\rho_r, \rho_d, \rho_s, \rho_0)$)
is also suggested by Theorem \ref{decrhor_incMK}.
The following heuristic argument suggests that in the massive MIMO regime, by increasing $M \propto 1/\sqrt{G_c}$ ($K$ fixed) the EE decreases at most by $10$ dB for every $20$ dB
reduction in channel gain. When $G_c$
reduces by $20$ dB, $M$ increases by $10$ dB, and therefore array gain increases by $10$ dB. This implies that the UTs need to increase their radiated power only by $20 - 10 = 10$ dB to maintain the same $R$.
Also, an increase in $M$ by a factor of $10$ increases the power consumption in the BS at most by a factor of $10$. Hence the total power consumption increases by a factor of at most $10$.

From the above discussion it is clear that by increasing $(M,K)$ with decreasing $G_c$ the
EE can be made to reduce slowly compared to a linear decrease with fixed $(M=2, K=1)$.
However, does the optimal EE decrease or can we vary $(M,K)$
in such a way that it actually increases with decreasing $G_c$? Theorem \ref{decGc_deceta}
suggests that the optimal EE always decreases with decreasing $G_c$. This conclusion
is indeed verified through exhaustive numerical simulations (see Fig.~\ref{rev_fig_1} in Section \ref{sec_sim}).

The other scenario is where $G_c$ is fixed and $(p_r, p_d, p_s, C_0)$ decrease
proportionately with a constant ratio between them.
With decreasing $(p_r, p_d, p_s, C_0)$ and fixed information rate, the power consumed by the BS and
the transmitter circuitry in the UT decreases whereas the power consumed by the PA remains unchanged
since $(R, G_c)$ is fixed. With decreasing $(p_r, p_d, p_s, C_0)$ and fixed $(M,K) = (2,1)$
the total system power consumption would be increasingly dominated by the power consumed by the PA at the UT.
Since the power consumed by the PA is fixed and it dominates the total power consumption,
it can be concluded that the EE with fixed $(M,K) = (2,1)$ would increase slowly
and approach a limit with decreasing $(p_r, p_d, p_s, C_0)$.
Can we increase the EE at a much faster rate by varying $(M,K)$ with
reducing $(p_r, p_d, p_s, C_0)$?

The answer is affirmative, With decreasing $(p_r, p_d, p_s, C_0)$, $(M,K)$ should be increased in a controlled manner such that both $(M p_r + K p_d + p_{mud})$ and $\alpha K p_u$ decrease, so that the
EE increases.
This is indeed possible. Corollary \ref{corr_9191} shows that the optimal EE increases strictly
with decreasing $(p_r,p_d,p_s,C_0)$. Theorem \ref{decrhor_incMK} also suggests that the optimal $(M,K)$
must be increased with decreasing $(p_r,p_d,p_s,C_0)$.
$\hfill\blacksquare$
\end{myremark}

\begin{myremark}
\label{remark_large_R}
From Remark \ref{remark_small_r} we know that with increasing channel gain $G_c$ and fixed $(\alpha, R , T, p_r, p_d, p_s, C_0)$, the optimal $K$ decreases. However, with decreasing $K$ and fixed $R$ the per-user information rate $R/K$
will increase, which is in turn expected to increase $p_d = p_{dec} + p_t$ (since the complexity of each user's channel decoder at the BS will increase due to the increased per-user information rate). If this increase in $p_d$ is significant, then it could nullify the expected increase in the optimal EE due to increasing channel gain. Note that this increase in $p_{dec}$ does not happen for sufficiently large $G_c$, since the optimal $K=1$ and therefore the per-user information rate is fixed (see Theorem \ref{thm6_stmt}).
For small $R$, irrespective of the value of $G_c$ this variation in $p_{dec}$ is expected to be small
compared to the value of $p_d$, since the variation in the per-user information rate is small.

For large $R$ and large $G_c$ the optimal $K$ could be small, leading to a very high
channel decoder complexity at the BS (e.g., $R= 50$ and an optimal $K^{\star}_{zf}(R, \Theta) = 2$ would result in a per-user information rate of $25$ bps/Hz).
In such scenarios one would prefer to have a higher $K$ instead of the optimal $K^{\star}_{zf}(R, \Theta)$ (e.g., with $K=16$ the per-user information rate reduces from $25$ bps/Hz to $50/16 = 3.125$ bps/Hz).
Reducing the per-user information rate would help in reducing $p_{dec}$ significantly.
However, with a sub-optimal $K > K^{\star}_{zf}(R, \Theta)$ the EE could decrease.
In the following theorem we show that by choosing a larger $K$,
the decrease in EE ${\zeta_{zf}^{\prime}(R, \Theta)}/{\zeta_{zf}^{\prime}(K, R, \Theta)}$ is upper bounded
by the ratio $2K/K^{\prime}_{csi}(R, \Theta)$. Here ${\zeta_{zf}^{\prime}(K, R, \Theta)}$ is the optimal
EE for a fixed $K$.
$\hfill\blacksquare$
\end{myremark}

\begin{mytheorem}
\label{thm_fixK}
Consider a $(R, \Theta)$ such that $\Theta$ satisfies (\ref{theta_cnds_1}) and $K_{csi}^{\prime}(R , \Theta) < 3K_{max}(\Theta)/4$.
Let the optimal EE with a fixed $K$ be given by
{
\vspace{-2mm}
\begin{eqnarray}
\label{eqn_fixK}
\zeta_{zf}^{\prime}(K, R, \Theta) & \Define & \hspace{-4mm} \max_{\substack{(M , \tau) \in {\mathbb R}^2 \,\vert \, \\
M \, > \, K \,\,,\,\, K \leq \tau < T }} \, \hspace{-3mm} \zeta_{zf}(M, K, \tau, R, \Theta).
\end{eqnarray}
}
Then for any $K$ satisfying $\frac{4}{3} K_{csi}^{\prime}(R , \Theta) < K < K_{max}(\Theta)$, the decrease in the EE (due to a suboptimal choice of $K$) satisfies
{
\vspace{-2mm}
\begin{eqnarray}
\label{bnd_eff_eqn}
1 & \leq \, \frac{\zeta_{zf}^{\prime}(R, \Theta)}{\zeta_{zf}^{\prime}(K, R, \Theta)}  \, < & \frac{2K}{K_{csi}^{\prime}(R, \Theta)}
\end{eqnarray}
}
\end{mytheorem}

{\em Proof}:
Refer to Appendix \ref{prf_thm_fixK}.
$\hfill\blacksquare$

\section{Simulation results}
\label{sec_sim}
\begin{figure}[t]
\begin{center}
\epsfig{file=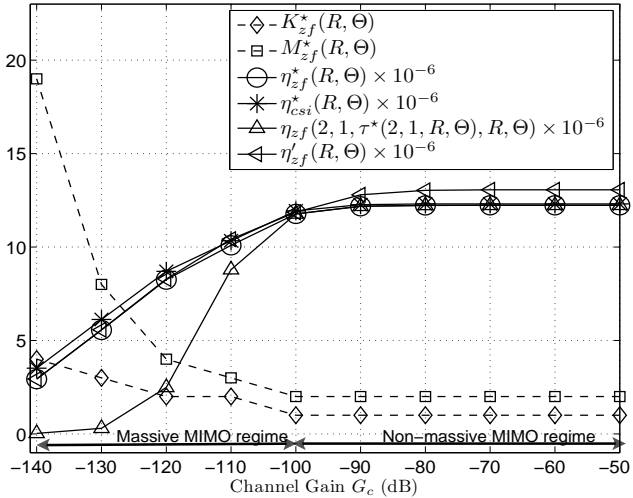, width=85mm,height=68mm}
\end{center}
\vspace{-3mm}
\caption{Unnormalized optimal EE $\eta_{zf}^{\star}(R, \Theta)$ versus channel gain $G_c$ for a fixed
$(\alpha, R, T, p_r, p_d, p_s, C_0, B, T_c)$. $\alpha =2$, $R = 8$ bps/Hz, $p_r = p_d = 0.01$ W, $p_s = 0.1$ W, $C_0 = 10^{-9}$ J, $B = 200$ KHz and $T_c = 2$ ms. $\eta_{zf}^{\star}(R, \Theta) = \eta_{zf}^{\prime\prime}(R, \Theta)$ since
$K^{\star}_{zf}(R, \Theta) < K_{max}(\Theta) = 16$ for the range of values considered for $G_c$.}
\label{rev_fig_1}
\vspace{-2mm}
\end{figure}
For all the numerical results in this section, we have taken $N_0 = 10^{-20.4}$ W/Hz, $T_c = 2$ ms, and $B = 200$ KHz.
In Fig.~\ref{rev_fig_1} we consider a fixed $p_r = p_d = 0.01$ W, $p_s = 0.1$ W,
$C_0 = 10^{-9}$ J, $\alpha =2$ and $R = 8$ bps/Hz. For these parameter values,
we get $K_{max}(\Theta) = 16$. We study the variation in the
unnormalized optimal EE $\eta^{\star}_{zf}(R, \Theta) = G_c \zeta^{\star}_{zf}(R, \Theta) / N_0$
as a function of decreasing channel gain $G_c$.
From the figure it is observed that for large $G_c > -100$ dB, $(M^{\star}_{zf}(R, \Theta) \,,\, K^{\star}_{zf}(R, \Theta)) = (2,1)$ (i.e., non-massive MIMO regime) as shown by Theorem \ref{thm6_stmt}.
From the analytical condition in (\ref{final_suff_cnd}) of Theorem \ref{thm6_stmt}, we get that $(M^{\star}_{zf}(R, \Theta) \,,\, K^{\star}_{zf}(R, \Theta)) = (2,1)$ for $G_c > -97$ dB, which agrees well with the observation from the figure. We also note from the figure that for large $G_c$ the optimal EE
remains almost constant with changing $G_c$. This supports the analytical observation from Theorem \ref{thm7_stmt}.
Refer to Remark \ref{remark_large_r} and Remark \ref{remark_large_r_1} for more discussion and insights.

With further decrease in $G_c$, it is observed that the optimal $(M,K)$ starts increasing,
as suggested by Theorem \ref{decrhor_incMK}. This regime is referred to as the massive MIMO regime.
For the range of values of $G_c$ considered in the figure, we observe that $K^{\star}_{zf}(R, \Theta) < K_{max}(\Theta)$ which implies that
the additional constraint $K \leq K_{max}(\Theta)$ does not have any impact on (\ref{prob_primeprime}) and therefore
$\eta_{zf}^{\star}(R,\Theta) = \eta_{zf}^{\prime\prime}(R, \Theta)  \Define  G_c \zeta_{zf}^{\prime\prime}(R, \Theta) / N_0$. We also plot the unnormalized optimal EE obtained by relaxing the integer constraints
on $(M,K,\tau)$, and observe that $\eta_{zf}^{\star}(R, \Theta) = \eta_{zf}^{\prime\prime}(R, \Theta) \approx \eta_{zf}^{\prime}(R, \Theta)  \Define  G_c \zeta_{zf}^{\prime}(R, \Theta) / N_0$ when $(M,K)$ are sufficiently large, i.e., the relaxation is tight as argued in the text following (\ref{prob_primeprime}). We also plot $\eta_{csi}^{\star}(R, \Theta) = G_c \zeta_{csi}^{\star}(R, \Theta)/N_0$ where $\zeta_{csi}^{\star}(R, \Theta)$ is the optimal EE of the ideal system
with $K \leq K_{max}(\Theta)$ and integer constraints on $(M,K)$. From Theorem \ref{thm2_stmt} it follows that
for $-141$ dB $< G_c < -96$ dB, the ratio $ \eta_{zf}^{\prime}(R,\Theta) / \eta_{csi}^{\prime}(R,\Theta)$ is bounded between $1$ and $3/8$. From numerical simulations we find that this is true even when $(M,K,\tau)$ are restricted
to be integers. In Fig.~\ref{rev_fig_1} $\eta_{zf}^{\star}(R,\Theta) \approx \eta_{csi}^{\star}(R, \Theta)$ for $-141$ dB $< G_c  < -96 $ dB.

From the figure we also observe that in the massive MIMO regime,
$\eta_{zf}^{\star}(R, \Theta)$ decreases with decreasing channel gain $G_c$. This confirms Theorem \ref{decGc_deceta}.
The same type of variation in the optimal EE of both the ideal and non-ideal systems (for scenarios where $p_{mud} \leq (M p_r + K p_d)$ and $K \leq K_{max}(\Theta) \leq T/4$), supports our hypothesis of studying the ideal system to make conclusions about the non-ideal system.

In Fig.~\ref{rev_fig_1} we also plot $\eta_{zf}(2,1,\tau^{\star}(2,1,R, \Theta), R , \Theta)$ which is the unnormalized
optimal EE for a fixed $(M,K) = (2,1)$ (i.e., $\zeta_{zf}(2,1,\tau,R,\Theta)$ is maximized over
integral values of $1 \leq \tau < T$). As discussed in Remark \ref{remark_small_r}, from the figure it is observed that for decreasing values of $G_c$ (when $M,K$ are larger than $2$ and $1$ respectively), the decrease in the optimal EE (with increasing $(M,K)$) is much smaller than the decrease for a fixed $(M,K) = (2,1)$.
As an example, when the channel gain decreases from $-100$ dB to $-130$ dB, the optimal EE with fixed $(M,K) = (2,1)$ decreases by a factor of about $39$, whereas the optimal EE with varying $(M,K)$
decreases by roughly $2$ times.
For a fixed $(M,K,R)$ it is clear that a $30$ dB decrease
in $G_c$ (from $-100$ to $-130$ dB) will increase the power consumed by the PAs
by a factor of roughly $10^3$. In contrast, for the case of varying $(M,K)$, using the optimal values of $(M,K)$
from the figure, we observe that at $G_c = -100$ dB $\alpha K p_u \approx  0.041$ W, whereas at $G_c = -130$ dB $\alpha K p_u \approx 0.051$ W, i.e., the total power consumed by the PAs increases by only
about $1.25$ times.
Additionally, the power radiated by each UT in fact decreases marginally from $0.02$ W at $G_c = -100$ dB to about $0.017$ W at $G_c = -130$ dB. Thus, varying $(M,K)$ in the massive MIMO regime helps to reduce the dynamic range requirement for the PAs in the UTs which would help
in improving PA linearity. Even in the non-massive MIMO regime i.e., when $G_c > -100$ dB, we could
keep the power radiated from the UT to be fixed at $0.02$ W and still achieve near-optimal EE (since in the non-massive MIMO regime, the power consumed by the PA is anyways significantly smaller than the power
consumed by other sources of power consumption, see Remark \ref{remark_large_r} and Remark \ref{remark_large_r_1}). The reduced dynamic range requirement for the PAs validates the linear PA model assumed by us in Section \ref{sec_pow_model}.
\begin{figure}[t]
\begin{center}
\epsfig{file=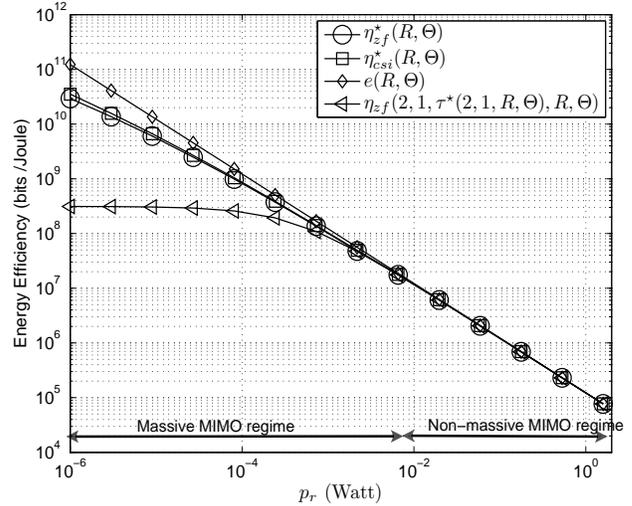, width=85mm,height=68mm}
\end{center}
\vspace{-3mm}
\caption{Unnormalized optimal EE $\eta_{zf}^{\star}(R, \Theta)$ versus $p_r$ for a fixed
$(\alpha, R, T, \beta, \delta, \mu, B, T_c, G_c)$. $\alpha =2$, $R = 8$ bps/Hz, $\beta = 1$, $\delta = 10$, $\mu = 0.02$, $B = 200$ KHz, $G_c = 10^{-10}$ (-$100$ dB) and $T_c = 2$ ms.}
\label{rev_fig_4}
\vspace{-2mm}
\end{figure}

In Fig.~\ref{rev_fig_4} we consider the scenario where $(\alpha, R, T_c, B, G_c)$ and the ratios between
the PCPs i.e., $(\beta, \delta, \mu)$ are fixed. To be precise,
$\alpha = 2$, $R = 8$ bps/Hz, $G_c = 10^{-10}$ ($-100$ dB), $\beta = 1$, $\delta = 10$ and $\mu = 0.02$.
We plot the unnormalized
optimal EE $\eta_{zf}^{\star}(R, \Theta)$ as a function of decreasing $p_r$. Such a scenario
models reduction in the PCPs due to technology scaling.
As shown in Theorem \ref{thm6_stmt}, it is observed from the figure that for large values of $p_r > 7.3 \times 10^{-3}$ W, $\eta_{zf}^{\star}(R, \Theta)$ increases linearly with decreasing $p_r$ (the slope of the log-log curve is $-1$). From the analytical condition in (\ref{final_suff_cnd}) of Theorem \ref{thm6_stmt}, we get that $(M^{\star}_{zf}(R, \Theta) \,,\, K^{\star}_{zf}(R, \Theta)) = (2,1)$ for $p_r > 5.5 \times 10^{-3}$ W (which agrees well with $p_r > 7.3 \times 10^{-3}$ W observed from the figure). We also observe that in this non-massive MIMO regime
the optimal EE is roughly the same as the bound $e(R, \Theta)$, which confirms Theorem \ref{thm7_stmt}.
The bound $e(R, \theta)$ assumes $p_u = 0$, and since it is tight, it follows that for large values of $(p_r,p_d, p_s, C_0)$
the power consumed by the PA in the UT is a small fraction of the total system power consumption (see Remark \ref{remark_large_r_1}).

With further decrease in $p_r$, the PCPs satisfy condition (\ref{theta_cnds_2}) when $8 \times 10^{-7} < p_r < 2.1 \times 10^{-2}$.
From Lemma \ref{lemma_918} in Appendix \ref{appendix_prev_work} we expect that the optimal $K_{zf}^{\star}(R, \Theta)$
is greater than one under these conditions (i.e., massive MIMO regime). 
The optimal EE increases with decreasing $p_r$, which confirms Corollary \ref{corr_9191}.
However, from the figure it is also observed that the rate of increase in the optimal EE
is less in the massive MIMO regime when compared to that in the non-massive MIMO regime. We believe this to be due to the increasing optimal $(M,K)$ in the massive MIMO regime.
In Fig.~\ref{rev_fig_4} we have also plotted the EE achieved with a fixed $(M,K) = (2,1)$.
As discussed in Remark \ref{remark_small_r}, it is observed that in the massive MIMO regime the EE can be improved
significantly by varying $(M,K)$ with changing $p_r$ as opposed to having a fixed $(M,K) = (2,1)$ (compare
the curves for $\eta_{zf}^{\star}(R, \Theta)$ and $\eta_{zf}(2,1,\tau^{\star}(2,1,R, \Theta),R,\Theta)$).
\begin{figure}[t]
\begin{center}
\epsfig{file=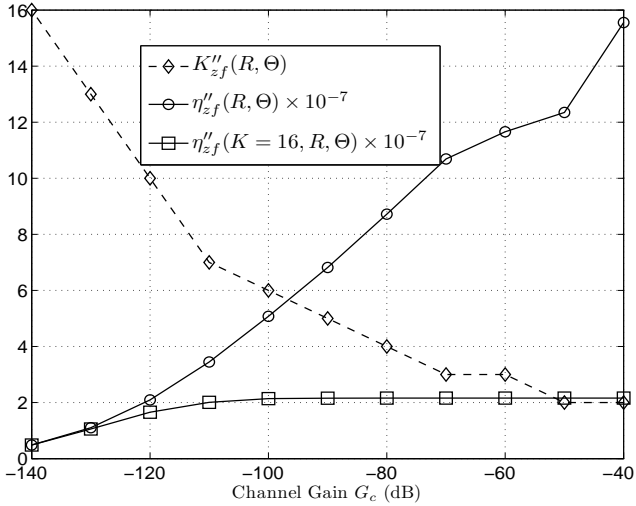, width=85mm,height=68mm}
\end{center}
\vspace{-3mm}
\caption{Optimal EE with a fixed $K=K_{max}(\Theta) = 16$, i.e., $\eta_{zf}^{\prime\prime}(K=16, R, \Theta)$ plotted as a function of varying channel gain, for a fixed high spectral efficiency $R=50$ bps/Hz. Fixed $\alpha = 2$,
$p_r = p_d = p_s = 0.01$ W, $C_0 = 10^{-9}$ J, $B = 200$ KHz and $T_c = 2$ ms.}
\label{rev_fig_5}
\vspace{-2mm}
\end{figure}

In Fig.~\ref{rev_fig_5}, we consider a fixed high
spectral efficiency of $R=50$ bps/Hz, where the channel decoder complexity and its power consumption (i.e., $p_{dec}$) would vary significantly due to varying per-user information rate.
Therefore for large $R$ it
it would not make practical
sense to have the optimal $K$ based on our model where $p_{dec}$ is assumed to be constant.
For such scenarios,
in Remark \ref{remark_large_R}, it is proposed that one should have the number of users to be significantly
larger than the optimal $K$ based on our system model. In Fig.~\ref{rev_fig_5} we plot the
EE achieved with a fixed $K = K_{max}(\Theta) = 16$ as a function of decreasing $G_c$, for a fixed $R = 50$ bps/Hz, $\alpha = 2$, $p_r = p_d = p_s = 0.01$ W, $C_0 = 10^{-9}$ J, $B = 200$ KHz and $T_c = 2$ ms.
The EE with a fixed $K$ is denoted by $\eta_{zf}^{\prime\prime}(K,R, \Theta)$, and is computed
by maximizing $\eta_{zf}(M, K, \tau, R , \Theta) = G_c \eta_{zf}(M, K, \tau, R , \Theta) / N_0$ jointly over $(M, \tau) \in {\mathbb Z}^2$
subject to $M > K$ and $K \leq \tau < T$.
In the figure,
at $G_c = -70$ dB, the per-user information rate with optimal number of users (based on our model, i.e., constant $p_d$)
is $R/K^{\prime\prime}_{zf}(R, \Theta) = 50/3$ bps/Hz. This cannot be a practical operating point due to the prohibitive
complexity of the channel decoders in the BS. Instead, with fixed $K = 16$ (curve marked with squares) the per-user information rate is only $50/16 = 3.125$ bps/Hz, and interestingly the EE is only $5$ times less than $\eta^{\prime\prime}_{zf}(R, \Theta)$ (as suggested by Theorem \ref{thm_fixK}).
\appendices
\section{Proof of Theorem \ref{thm91_stmt}}
\label{prf_thm91_stmt}
The proof follows from the following where we have used the abbreviations $M^{\star}_1, K^{\star}_1, \tau^{\star}_1$
for $M_{zf}^{\star}(R , \Theta_1), K_{zf}^{\star}(R , \Theta_1), \tau^{\star}(R , \Theta_1)$ respectively,
and similarly $M^{\star}_2, K^{\star}_2, \tau^{\star}_2$
for $M_{zf}^{\star}(R , \Theta_2), K_{zf}^{\star}(R , \Theta_2), \tau^{\star}(R , \Theta_2)$ respectively.
{
\small
\vspace{-2mm}
\begin{eqnarray}
\frac{R}{\zeta_{zf}^{\star}(R , \Theta_1)} & {(a) \atop =}  & \frac{R}{\zeta_{zf}(M^{\star}_1, K^{\star}_1, \tau^{\star}_1, R , \Theta_1) }    \nonumber \\
&  {(b) \atop \leq }  &  \frac{R}{\zeta_{zf}(M^{\star}_2, K^{\star}_2, \tau^{\star}_2, R , \Theta_1) } \nonumber \\
& \hspace{-24mm} {(c) \atop =}  & \hspace{-15mm} {\Bigg (} \alpha K^{\star}_2 \gamma^{\star}_2 \, + \, \rho_{s_1}  \, + \, K^{\star}_2 {\Big (} \rho_{d_1} \, + \, \frac{8}{3} {(K^{\star}_2)}^2  \frac{\rho_{0_1}}{T} {\Big )}   \nonumber \\
& &  + M^{\star}_2 {\Big (} \rho_{r_1} \, + \, 2 K^{\star}_2 \rho_{0_1} \, + \, 4 {(K^{\star}_2)}^2 \frac{\rho_{0_1}}{T}{\Big )}  \, {\Bigg )}\nonumber \\
& \hspace{-24mm} {(d) \atop  < } & \hspace{-15mm} {\Bigg (} \alpha K^{\star}_2 \gamma^{\star}_2 \, + \, \rho_{s_2} \, + \, K^{\star}_2 {\Big (} \rho_{d_2} \, + \, \frac{8}{3} {(K^{\star}_2)}^2  \frac{\rho_{0_2}}{T} {\Big )}  \nonumber \\
& &  + M^{\star}_2 {\Big (} \rho_{r_2} \, + \, 2 K^{\star}_2 \rho_{0_2} \, + \, 4 {(K^{\star}_2)}^2 \frac{\rho_{0_2}}{T}{\Big )} \, {\Bigg )}  \nonumber \\ 
& \hspace{-24mm} = & \hspace{-15mm} \frac{R}{\zeta_{zf}^{\star}(R , \Theta_2)}
\end{eqnarray}
\normalsize
}
where (a) is due to (\ref{opt_int_M_K_tau}), (b) follows
from (\ref{opt_int_M_K_tau}) being a minimization problem and
(c) follows from (\ref{zeta_zf_1}). For step (d) note
that at least one among $(\rho_{r_1}, \rho_{d_1}, \rho_{s_1}, \rho_{0_1})$ is strictly less than
its counterpart in $\Theta_2$.
$\hfill\blacksquare$

\section{Proof of Theorem \ref{thm6_stmt}}
\label{prf_thm6_stmt}
Consider the set
{
\vspace{-2mm}
\begin{eqnarray}
{\mathcal M} & \Define & {\Bigg \{}  (M,K) \in {\mathbb Z}^2 \,\vert\,  M > K \geq 1 \,,\, (M,K) \ne (2,1) {\Bigg \}}.  \nonumber \\
\end{eqnarray}
}
{
\vspace{-7mm}
\begin{eqnarray}
\frac{R}{\zeta_{zf}^{\star}(R , \Theta)} & \hspace{-3mm}  = &  \hspace{-3mm} \min{\Bigg (} \min_{\substack{(M,K) \in {\mathcal M} \,,\, \tau \in {\mathbb Z}, \\
K \leq \tau < T}} \, \frac{R}{\zeta_{zf}(M , K, \tau , R , \Theta) }   \nonumber \\
 & & \hspace{-2mm} , \,  \min_{\substack{\tau \in {\mathbb Z} \,,\, \\ 1 \leq \tau < T}}  \, \frac{R}{\zeta_{zf}(M=2,K=1,\tau, R , \Theta) } {\Bigg )}
\end{eqnarray}
}
and therefore in order to show that $(M_{zf}^{\star}(R, \Theta) \,,\, K_{zf}^{\star}(R , \Theta)) = (2,1)$
it suffices to show
{
\vspace{-3mm}
\begin{eqnarray}
\label{eqn_46}
\min_{\substack{(M,K) \in {\mathcal M} \,,\, \tau \in {\mathbb Z}, \\
K \leq \tau < T}} \, \frac{R}{\zeta_{zf}(M , K, \tau , R , \Theta) }   \nonumber \\
&  & \hspace{-23mm} > \, \min_{\substack{\tau \in {\mathbb Z} \,,\, \\ 1 \leq \tau < T}} \, \frac{R}{\zeta_{zf}(2,1,\tau, R , \Theta)}.
\end{eqnarray}
}
Since
$\frac{R}{\zeta_{zf}(2,1,\lfloor \sqrt{T} \rfloor, R , \Theta)} \, >  \, \min_{\substack{\tau \in {\mathbb Z} \,,\, \\ 1 \leq \tau < T}} \, \frac{R}{\zeta_{zf}(2,1,\tau, R , \Theta)}$
a sufficient condition which guarantees (\ref{eqn_46}) is
that, for all $(M,K, \tau)$ such that $(M,K) \in {\mathcal M}$ and $K \leq \tau < T$
{
\vspace{-1mm}
\begin{eqnarray}
\label{eqn_47}
\frac{R}{\zeta_{zf}(M, K, \tau, R , \Theta)} & \hspace{-3mm} > & \hspace{-3mm} \frac{R}{\zeta_{zf}(2,1,\lfloor \sqrt{T} \rfloor, R , \Theta)}.
\end{eqnarray}
}
Using the upper bound on $\gamma_u$ in (\ref{eqn_22}) of Lemma \ref{lemma_gammau_ub} with $(M,K,\tau) = (2,1,\lfloor \sqrt{T} \rfloor)$ we get
{
\vspace{-3mm}
\begin{eqnarray}
\label{eqn_73}
\frac{R}{\zeta_{zf}(2,1,\lfloor \sqrt{T} \rfloor, R , \Theta)} & \hspace{-3mm} = &  \hspace{-3mm}  \alpha \gamma_u + \rho_s
+ 2\rho_r + \rho_d + 4 \rho_0 + \frac{32}{3} \frac{\rho_0}{T} \nonumber \\
& \hspace{-36mm}  < & \hspace{-18mm} {\Bigg [} \frac{\alpha}{1 + \lfloor \sqrt{T} \rfloor} \, + \, \frac{\alpha (1 + \lfloor \sqrt{T} \rfloor)}{\lfloor \sqrt{T} \rfloor}
{\Big (} 2^{\frac{R}{1 - \frac{\lfloor \sqrt{T} \rfloor}{T}}} - 1 {\Big )}   \nonumber \\
& & \hspace{-5mm} + \rho_s + 2\rho_r + \rho_d + 4 \rho_0 + \frac{32}{3} \frac{\rho_0}{T} {\Bigg ]}.
\end{eqnarray}
}
Also, for any $(M, K, \tau)$ from (\ref{zeta_zf_1}) it is clear that
{
\vspace{-3mm}
\begin{eqnarray}
\label{eqn_74}
\frac{R}{\zeta_{zf}(M,K,\tau, R, \Theta)} & > & M {\Big (} \rho_r \, + \, 2 K \rho_0 \, + \, 4 K^2 \frac{\rho_0}{T}{\Big )}   \nonumber \\
& +  &  K {\Big (} \rho_d \, + \, \frac{8}{3} K^2  \frac{\rho_0}{T} {\Big )} + \rho_s.
\end{eqnarray}
}
A sufficiency condition for (\ref{eqn_47}) (and therefore for (\ref{eqn_46})) is that the R.H.S. of (\ref{eqn_74})
is greater than the R.H.S. of (\ref{eqn_73}) for all $(M,K) \in {\mathcal M}$, i.e.
{
\vspace{-2mm}
\begin{eqnarray}
\label{eqn_57}
M {\Big (} \rho_r \, + \, 2 K \rho_0 \, + \, 4 K^2 \frac{\rho_0}{T}{\Big )} +   K {\Big (} \rho_d \, + \, \frac{8}{3} K^2  \frac{\rho_0}{T} {\Big )} + \rho_s & & \nonumber \\
& & \hspace{-80mm} \, > \, {\Bigg [} \frac{\alpha}{1 + \lfloor \sqrt{T} \rfloor} \, + \, \frac{\alpha (1 + \lfloor \sqrt{T} \rfloor)}{\lfloor \sqrt{T} \rfloor}
{\Big (} 2^{\frac{R}{1 - \frac{\lfloor \sqrt{T} \rfloor}{T}}} - 1 {\Big )}   \nonumber \\
&  &  \hspace{-60mm} + \rho_s + 2\rho_r + \rho_d + 4 \rho_0 + \frac{32}{3} \frac{\rho_0}{T} {\Bigg ]}.
\end{eqnarray}
}
For any $(M,K) \in {\mathcal M}$ it is clear that
{
\vspace{-2mm}
\begin{eqnarray}
\label{eqn_69}
M {\Big (} \rho_r \, + \, 2 K \rho_0 \, + \, 4 K^2 \frac{\rho_0}{T}{\Big )}
 +  K {\Big (} \rho_d \, + \, \frac{8}{3} K^2  \frac{\rho_0}{T} {\Big )} + \rho_s   \nonumber \\
&  &  \hspace{-80mm} > \, 3 {\Big (} \rho_r + 2 \rho_0 + 4 \frac{\rho_0}{T} {\Big )} \, + \, {\Big (} \rho_d + \frac{8}{3} \frac{\rho_0}{T}{\Big )} \, + \, \rho_s \nonumber \\
&  &  \hspace{-80mm}  >  \, \rho_s + 3 \rho_r + \rho_d + 6 \rho_0 + \frac{32}{3} \frac{\rho_0}{T}.
\end{eqnarray}
}
Using (\ref{eqn_69}) in (\ref{eqn_57}) we see that (\ref{eqn_57}) is guaranteed if
{
\small
\vspace{-1mm}
\begin{eqnarray}
\label{eqn_162_p}
\rho_s + 3 \rho_r + \rho_d + 6 \rho_0 + \frac{32}{3} \frac{\rho_0}{T} & > & {\Bigg [} \rho_s + 2 \rho_r + \rho_d + 4 \rho_0  \nonumber \\
& & \hspace{-49mm}  + \frac{32}{3} \frac{\rho_0}{T} 
 +  \frac{\alpha}{1 + \lfloor \sqrt{T} \rfloor} \, + \, \frac{\alpha (1 + \lfloor \sqrt{T} \rfloor)}{\lfloor \sqrt{T} \rfloor}
{\Big (} 2^{\frac{R}{1 - \frac{\lfloor \sqrt{T} \rfloor}{T}}} - 1 {\Big )} {\Bigg ]}
\end{eqnarray}
\normalsize
}
which completes the proof.
$\hfill\blacksquare$

\section{Proof of Theorem \ref{thm7_stmt}}
\label{prf_thm7_stmt}
From Theorem \ref{thm6_stmt} we know that, for any $(R,\Theta)$ satisfying the conditions in
(\ref{final_suff_cnd}), the optimal $(M,K) = (2,1)$ and therefore
{
\vspace{-3mm}
\begin{eqnarray}
\label{zeta_21_lb}
\frac{R}{\zeta_{zf}^{\star}(R, \Theta)} & = & \min_{\substack{\tau \in {\mathbb Z} \,,\, \\ 1 \leq \tau < T}} \, \frac{R}{\zeta_{zf}(2 , 1, \tau , R , \Theta)}   \nonumber \\
& {(a) \atop > }  & \rho_s \, + \, 2 \rho_r \, + \, 4 \rho_0 \, + \, \rho_d \, + \, \frac{32}{3} \frac{\rho_0}{T}
\end{eqnarray}
}
where (a) follows from the fact that $\gamma_u > 0$ for any $1 \leq \tau < T$.
Again using Theorem \ref{thm6_stmt}, for any $(R, \Theta)$ satisfying the conditions in
(\ref{final_suff_cnd})
we also have
{
\vspace{-2mm}
\begin{eqnarray}
\label{zeta_21_ub}
\frac{R}{\zeta_{zf}^{\star}(R, \Theta)} & = & \min_{1 \leq \tau < T} \, \frac{R}{\zeta_{zf}(2 , 1, \tau , R , \Theta)}   \nonumber \\
&  \hspace{-28mm} {(a) \atop =}  &   \hspace{-14mm} \alpha \, {\Big (} \min_{1 \leq \tau < T} \, \gamma_u  {\Big )} \, +  \, \rho_s \, + \, 2 \rho_r \, + \, 4 \rho_0 \, + \, \rho_d \, + \, \frac{32}{3} \frac{\rho_0}{T} \nonumber \\
& \hspace{-28mm} {(b) \atop < } & \hspace{-14mm} \alpha \, \gamma_u(\tau = \lfloor \sqrt{T} \rfloor) \, + \, \rho_s \, + \, 2 \rho_r \, + \, 4 \rho_0 \, + \, \rho_d \, + \, \frac{32}{3} \frac{\rho_0}{T} \nonumber \\
& \hspace{-28mm}  { (c) \atop < } & \hspace{-14mm} \frac{\alpha}{1 + \lfloor \sqrt{T} \rfloor} \, + \, \frac{\alpha (1 + \lfloor \sqrt{T} \rfloor)}{\lfloor \sqrt{T} \rfloor}
{\Big (} 2^{\frac{R}{1 - \frac{\lfloor \sqrt{T} \rfloor}{T}}} - 1 {\Big )}   \nonumber \\
& &  + \, \rho_s \, + \, 2 \rho_r \, + \, 4 \rho_0 \, + \, \rho_d \, + \, \frac{32}{3} \frac{\rho_0}{T} \nonumber \\
& \hspace{-28mm}  { (d) \atop \leq } &  \hspace{-14mm}  \rho_s \, + \, 3 \rho_r \, + \, 6 \rho_0 \, + \, \rho_d \, + \, \frac{32}{3} \frac{\rho_0}{T}   \nonumber \\
& \hspace{-28mm}  < &  \hspace{-14mm} \,  \frac{3}{2} {\Big (} \rho_s \, + \, 2 \rho_r \, + \, 4 \rho_0 \, + \, \rho_d \, + \, \frac{32}{3} \frac{\rho_0}{T}  {\Big )}
\end{eqnarray}
}
where (a) follows from the fact that in the expression for $\zeta_{zf}(2 , 1, \tau, R , \Theta)$
only $\gamma_u$ depends on $\tau$. Step (b) follows from the fact that $\tau = \lfloor \sqrt{T} \rfloor$ satisfies the condition
$1 \leq \tau < T$ in the minimization in step (a), since $T > 1$.
The notation $\gamma_u(\tau = \lfloor \sqrt{T} \rfloor)$ is used to highlight the fact that
we choose $\tau = \lfloor \sqrt{T} \rfloor$.
Step (c) follows from (\ref{eqn_22}) of Lemma \ref{lemma_gammau_ub} (with $(M,K,\tau) = (2,1,\lfloor \sqrt{T} \rfloor)$).
Step (d) follows from the fact that $(R, \Theta)$ satisfies (\ref{final_suff_cnd}).
Using (\ref{zeta_21_lb}) and (\ref{zeta_21_ub}) along with (\ref{rho_def}) and (\ref{eta_star_def})
completes the proof.
$\hfill\blacksquare$

\section{Proof of Lemma \ref{lemma_theta_cnds}}
\label{prf_lemma_theta_cnds}
From (\ref{mud_pow}) we know that
$p_{mud}$ is a sum of two terms, $2 M K C_0 B  +  (4 M K^2 {C_0}/{T_c})$ and $8 K^3 {C_0}/{(3 T_c)}$. We will show that if $\Theta$ satisfies the conditions in (\ref{theta_cnds}), then for all
$1 \leq K \leq K_{max}(\Theta)$ and any $M$, the first term of $p_{mud}$ is less than $M p_r$ and the second term is less
than $K p_d$.
The fact that the first term is less than or equal to $M p_r$ follows from the following chain of inequalities
\begin{eqnarray}
\label{first_term_eqn}
\frac{(2 M K C_0 B \, + \, 4 M K^2 \frac{C_0}{T_c})}{M p_r } & = & 2 K \frac{C_0 B}{p_r} \, + \, 4 K^2 \frac{C_0 }{p_r T_c}  \nonumber \\
& \hspace{-12mm} {(a) \atop = } & \hspace{-7mm} \, 2 K \frac{\rho_0}{\rho_r}  \, + \, 4 K^2 \frac{\rho_0}{T \rho_r} \nonumber \\
& \hspace{-12mm} {(b) \atop \leq } & \hspace{-7mm} \frac{ K_{max}(\Theta) \rho_0}{\rho_r} {\Big (} 2 \, + \, \frac{K_{max}(\Theta)}{T/4} {\Big )}   \nonumber \\
& \hspace{-12mm} {(c) \atop \leq } & \hspace{-7mm} \, 3 \frac{ K_{max}(\Theta) \rho_0}{\rho_r} \, \leq \, 1
\end{eqnarray}
where (a) follows from (\ref{rho_def}) and (b) follows from the fact that $K \leq K_{max}(\Theta)$.
Step (c) follows from (\ref{kmax_def}).
Next we show that the second term of $p_{mud}$ is less than or equal to $K p_d$.
\begin{eqnarray}
\label{second_term_eqn}
\frac{8 K^3 \frac{C_0}{3 T_c}}{K p_d} & { (a) \atop = } & \frac{ 2 K \rho_0}{3 \rho_d} \, \frac{4 K}{T}   \nonumber \\
& {(b) \atop \leq }  & \frac{ 2 K_{max}(\Theta) \rho_0}{3 \rho_d} \, \frac{4 K_{max}(\Theta)}{T}   \nonumber \\
& {(c) \atop \leq }  &  \, 1
\end{eqnarray}
where (a) follows from (\ref{rho_def}) and (b) follows from the fact that $K \leq K_{max}(\Theta)$.
Step (c) follows from (\ref{kmax_def}).
The inequality in (\ref{pmud_dom}) now follows from (\ref{first_term_eqn}) and (\ref{second_term_eqn}).
Also, from (\ref{kmax_def}) and condition (C.1) in (\ref{theta_cnds}) it follows
that $K_{max}(\Theta) > 10$.
$\hfill\blacksquare$

\section{Proof of Lemma \ref{lemma_37}}
\label{prf_lemma_37}
The equivalence between (\ref{theta_cnds}) and ((\ref{theta_cnds_1}), (D.1) of (\ref{theta_cnds_2}))
follows from (\ref{fixed_ratio}).
Using the definition of $c(\Theta)$ in (\ref{suff_cnd}), we observe that
(D.2) of (\ref{theta_cnds_2}) is equivalent to
\begin{eqnarray}
\label{eqn_651}
\frac{g(4R/3K_{max}(\Theta))}{\sqrt{4R / 3K_{max}(\Theta)}} & < & \frac{g(c(\Theta))}{\sqrt{c(\Theta)}}
\end{eqnarray}
Since $g(x)/\sqrt{x}$ is a strictly monotonically increasing with $x > 0$, it follows that
(\ref{eqn_651}) is equivalent to
$\frac{4}{3} \, \frac{R}{K_{max}(\Theta)}  <  c(\Theta)$, i.e.,
$R \,  <  \, \frac{3}{4} c(\Theta) K_{max}(\Theta) \, = \, \frac{3}{4} R_{max}(\Theta)$,
which completes the proof.
The uniqueness of $c(\Theta)$ in (\ref{suff_cnd})
is due to the fact that $g(x)/\sqrt{x}$ is unbounded
and strictly monotonically increasing.
$\hfill\blacksquare$

\section{Tight bounds on $\zeta_{zf}^{\prime}(R ,\Theta)$ for $(R , \Theta)$ satisfying (\ref{theta_cnds_1}) and (\ref{theta_cnds_2})}
\label{appendix_prev_work}
The following theorem presents the exact solution to (\ref{zf_csi_opt_MK}).
\begin{mytheorem}
\label{thm1_stmt}
For a given $(R , \Theta)$ there exists a unique $x^{\prime}(R , \Theta) \geq 0$ such that
{
\hspace{-1mm}
\begin{eqnarray}
\label{gxprime_eqn_rev}
g(x^{\prime}(R , \Theta)) & = & {\Big (} 1 + \frac{\rho_d}{{\rho_r}} {\Big ) } \, \sqrt{\frac{R \rho_r}{\alpha}}
\end{eqnarray}
}
where $g(x) \,,\, x \geq 0$ is given by (\ref{gx_def_eqn}).
For any given $\Theta$ satisfying (\ref{theta_cnds_1}), the solution to (\ref{zf_csi_opt_MK}) is given by
{
\vspace{-2mm}
\begin{eqnarray}
\label{opt_M_opt_K}
K_{csi}^{\prime}(R , \Theta) & = & \max {\Big (}  \, \min {\Big (} \frac{R}{x^{\prime}(R , \Theta)} \,,\, K_{max}(\Theta) {\Big )} \,,\, 1 {\Big )} \nonumber \\
M_{csi}^{\prime}(R , \Theta) & = & K_{csi}^{\prime}(R , \Theta)    \nonumber \\
& & \hspace{-7mm}  + \, \sqrt{K_{csi}^{\prime}(R , \Theta)} \sqrt{ \frac{\alpha \, (2^{R / K_{csi}^{\prime}(R , \Theta)} - 1)}{\rho_r}}.
\end{eqnarray}
}
\end{mytheorem}

{\em Proof:}
In (\ref{zf_csi_opt_MK}), in order to minimize $\frac{1}{\zeta_{csi}(M, K , R , \Theta)}$ w.r.t. both $M$ and $K$, we first minimize it w.r.t. $M$ for a given $(K , R , \Theta)$, i.e.,
{
\vspace{-4mm}
\begin{eqnarray}
\label{opt_M_given_K}
M^{\prime}_{csi}(K, R , \Theta)  & = & \arg \min_{M \in {\mathbb R} \,,\, M > K} \frac{1}{\zeta_{csi}(M, K , R ,  \Theta)}  \nonumber \\
& \hspace{-12mm} = &  \hspace{-9mm} \arg \min_{M \in {\mathbb R} \,,\, M > K}  \frac{K}{R} {\Big (} \alpha  \frac{2^{R/K} - 1}{M - K} + \rho_d + \frac{M}{K} \rho_r {\Big )} \nonumber \\
& \hspace{-12mm}  = & \hspace{-9mm}  K + \sqrt{\frac{\alpha (2^{R/K} - 1)}{\rho_r}} \sqrt{K}.
\end{eqnarray}
}
Using this in (\ref{eqn_332}) we get
\begin{eqnarray}
\label{zeta_zf_K}
\frac{1}{\zeta_{csi}^{\prime}( K, R , \Theta)} & \Define & \frac{1}{\zeta_{csi}( M^{\prime}_{csi}(K, R , \Theta) , K, R , \Theta)}   \nonumber \\
& \hspace{-42mm}  = & \hspace{-25mm }  \frac{1}{R} {\Bigg (} K ( \rho_d + \rho_r)  + 2 \sqrt{\alpha \rho_r K  (2^{R/K} - 1) } \, + \, \rho_s {\Bigg )}.
\end{eqnarray}
We next minimize $1/\zeta_{csi}^{\prime}( K, R , \Theta)$ w.r.t. $K$, subject to the constraint that $1 \leq K \leq K_{max}(\Theta)$. However we firstly consider finding the minimum of $1/\zeta_{csi}^{\prime}( K, R , \Theta)$ when $K$ is unconstrained and then introduce the constraints later. The unconstrained minimum is obtained by setting the derivative of $1/\zeta_{csi}^{\prime}( K, R , \Theta)$ w.r.t. $K$ to be zero.
Doing this, we see that the optimal $K^{\prime}$ must satisfy
\begin{eqnarray}
\label{deriv_zero_rev}
\rho_r + \rho_d + \sqrt{\frac{\alpha \rho_r (2^{R/K^{\prime}} - 1)} {K^{\prime}}} \, - \, \frac{2^{R/K^{\prime}} \frac{R}{K^{\prime}} \log(2) \alpha \rho_r}{\sqrt{\alpha \rho_r K^{\prime} (2^{R/K^{\prime}} - 1)}}  & \hspace{-3mm} = & \hspace{-3mm} 0. \nonumber \\
\end{eqnarray}
Further for any $(R, \Theta)$, the second derivative of $1/\zeta_{csi}^{\prime}( K, R , \Theta)$ w.r.t. $K$ is positive, i.e.
\begin{eqnarray}
\label{convex_1byzeta}
\frac{d^2 (1/\zeta_{csi}^{\prime}( K, R , \Theta))}{d K^2} > 0 \,\,\,,\,\,\, \forall K > 0
\end{eqnarray}
i.e., $1/\zeta_{csi}^{\prime}( K, R , \Theta)$ is a strictly convex function of $K > 0$ for any given $(R, \Theta)$.
Therefore
if there exists a solution to (\ref{deriv_zero_rev}), then it has to be the unique global unconstrained minimum
of $1/\zeta_{csi}^{\prime}( K, R , \Theta)$ w.r.t. $K$.
With the notation $r \Define R/K^{\prime}$, the condition in (\ref{deriv_zero_rev}) can be equivalently written as
{
\vspace{-1mm}
\begin{eqnarray}
\label{g_r_eqn_rev}
{\Big (} \sqrt{\rho_r} + \frac{\rho_d}{\sqrt{\rho_r}} {\Big )} \sqrt{\frac{R}{\alpha}} & = & g(r)
\end{eqnarray}
}
where $g(\cdot)$ is given by (\ref{gx_def_eqn}).
It can be shown that $g(r)$ is a strictly monotonically increasing function of $r$, $g(0) = 0$ and $d^2 g(r) / dr^2 > 0$. Since $(\sqrt{\rho_r} + \rho_d/\sqrt{\rho_r}) \sqrt{R/\alpha} \, > 0$, it follows that the solution\footnote{\footnotesize{``Solution to  (\ref{g_r_eqn_rev})'' refers to the value of $r > 0$ such that $g(r) = (\sqrt{\rho_r} + \rho_d/\sqrt{\rho_r}) \sqrt{R/\alpha}$.}} to (\ref{g_r_eqn_rev}) exists and is unique, and we denote it by $r = x^{\prime}(R , \Theta)$ (this proves (\ref{gxprime_eqn_rev})). Since $r = R/K^{\prime} = x^{\prime}(R , \Theta)$ is the unique solution to (\ref{g_r_eqn_rev}), it follows that the unique solution to (\ref{deriv_zero_rev}) is $K^{\prime} = R/x^{\prime}(R , \Theta)$.

Given that $R/x^{\prime}(R, \Theta)$ is the location of the unique minimum of the objective function $1/\zeta_{csi}^{\prime}( K, R , \Theta)$ without any constraints on $K$, we will next find the expression for the unique minimum of $1/\zeta_{csi}^{\prime}( K, R , \Theta)$ subject to the constraint $1 \leq K \leq K_{max}(\Theta)$.
If $R/x^{\prime}(R, \Theta)$ lies in the interval $[1 \,,\, K_{max}(\Theta)]$ then it is clear that $R/x^{\prime}(R, \Theta)$
will remain to be the unique minimum of the objective function $1/\zeta_{csi}^{\prime}( K, R , \Theta)$ under the constraint $1 \leq K \leq K_{max}(\Theta)$.
If the unconstrained minimum $R/x^{\prime}(R, \Theta) < 1$, then since the objective function is strictly convex in $K$ (from (\ref{convex_1byzeta})) it follows
that its derivative w.r.t. $K$ is strictly positive for all $K \in [1 \,,\, K_{max}(\Theta)]$ (the derivative is an increasing function of $K$ and is zero at $K = R/x^{\prime}(R, \Theta) < 1$).
Hence for the case when
$R/x^{\prime}(R, \Theta) < 1$, the unique minimum of the objective function in the interval $[1 \,,\, K_{max}(\Theta)]$ will be at $K^{\prime}_{csi}(R , \Theta) = 1$. Lastly, if $R/x^{\prime}(R, \Theta) > K_{max}(\Theta)$, then since the objective function is strictly convex in $K$ it follows
that its derivative w.r.t. $K$ is strictly negative for all $K \in [1 \,,\, K_{max}(\Theta)]$. Hence for the case when
$R/x^{\prime}(R, \Theta) > K_{max}(\Theta)$, the unique minimum of the objective function in the interval $[1 \,,\, K_{max}(\Theta)]$ will be at $K^{\prime}_{csi}(R , \Theta) = K_{max}(\Theta)$.
Combining all these cases we get the expression for $K^{\prime}_{csi}(R, \Theta)$ in (\ref{opt_M_opt_K}).
Further, from (\ref{opt_M_given_K}) it follows that the optimal $M$ is $M^{\prime}_{csi}(R , \Theta) = M^{\prime}_{csi}(K^{\prime}_{csi}(R, \Theta), R , \Theta)$.
$\hfill\blacksquare$

From Lemma \ref{lemma_37} we know that any $(R, \Theta)$ satisfies (D.2) of (\ref{theta_cnds_2})
if and only if $R < 3 R_{max}(\Theta)/4$.
Along with this fact, the following lemma shows that for any $(R , \Theta)$ satisfying (\ref{theta_cnds_1}) and
(\ref{theta_cnds_2}), $K_{csi}^{\prime}(R , \Theta)$ lies strictly between $1$ and
$K_{max}(\Theta)$. This result is useful later in deriving tight bounds on $\zeta_{zf}^{\prime}(R, \Theta)$.
\begin{mylemma}
\label{lemma_918}
For any $\Theta$ satisfying (\ref{theta_cnds_1}) and any $R$, we have $R < R_{max}(\Theta)$ if and only if
$R/x^{\prime}(R,\Theta) < K_{max}(\Theta)$, i.e.
{
\vspace{-2mm}
\begin{eqnarray}
\label{eqn_1231}
R \, < \, R_{max}(\Theta) & \Longleftrightarrow & \frac{R}{x^{\prime}(R, \Theta)} \, < \, K_{max}(\Theta).
\end{eqnarray}
}
or equivalently
{
\vspace{-2mm}
\begin{eqnarray}
\label{kprime_eqiv}
R \, < \, R_{max}(\Theta) & \Longleftrightarrow & K_{csi}^{\prime}(R , \Theta) \, < \, K_{max}(\Theta).
\end{eqnarray}
}
Further for any $\Theta$ satisfying (\ref{theta_cnds_1}), $(R, \Theta)$ satisfies (D.3) of (\ref{theta_cnds_2}) if and only if
$K_{csi}^{\prime}(R , \Theta)  >   1$, i.e.
{
\vspace{-1mm}
\begin{eqnarray}
\label{lemma_w2}
K_{csi}^{\prime}(R , \Theta)  & >  & 1.
\end{eqnarray}
}
\end{mylemma}

{\em Proof}:
Note that $R/x^{\prime}(R, \Theta) < K_{max}(\Theta)$ if and only if
$x^{\prime}(R, \Theta) > R/K_{max}(\Theta)$ which in turn holds if and only
if
{
\vspace{-1mm}
\begin{eqnarray}
g(x^{\prime}(R , \Theta)) &  \hspace{-5mm} { (a) \atop > } &  \hspace{-5mm} g{\Big (} {R}/{K_{max}(\Theta)} {\Big )} \,\,,\,\, \mbox{or equiv.}  \nonumber \\
 {\Big (} 1 + \frac{\rho_d}{\rho_r} {\Big )} \sqrt{\frac{R \rho_r}{\alpha}} \,\,\, &  { (b) \atop > } & \,\,\, g{\Big (} \frac{R}{K_{max}(\Theta)} {\Big )} \,\,\,,\,\,\, \mbox{or equiv.} \nonumber \\
{\Big (} 1 + \frac{\rho_d}{\rho_r} {\Big )} \sqrt{\frac{K_{max}(\Theta) \rho_r}{\alpha}} & > & \frac{g(R/K_{max}(\Theta) )}{\sqrt{R/K_{max}(\Theta)}}  \,\,\,,\,\,\, \mbox{or equiv.}  \nonumber \\
\frac{g(c(\Theta))}{\sqrt{c(\Theta)}} \,\,\,  & { (c) \atop > } & \,\,\, \frac{g(R/K_{max}(\Theta) )}{\sqrt{R/K_{max}(\Theta)}}
\end{eqnarray}
}
where (a) follows from the fact that $g(x)$ is strictly monotonically increasing with $x > 0$.
Step (b) follows from (\ref{gxprime_eqn_rev}) and step (c) follows from the definition of $c(\Theta)$ in (\ref{suff_cnd}).
Since $g(x)/\sqrt{x}$ is strictly monotonically increasing with $x > 0$, step (c) above is equivalent to
$c(\Theta) > R/K_{max}(\Theta)$, which is in turn equivalent to $R < c(\Theta) K_{max}(\Theta) = R_{max}(\Theta)$. This proves (\ref{eqn_1231}).

From the expression for $K^{\prime}_{csi}(R, \Theta)$ in (\ref{opt_M_opt_K}) it follows that
$R/x^{\prime}(R, \Theta) < K_{max}(\Theta)$ if and only if $K^{\prime}_{csi}(R, \Theta) < K_{max}(\Theta)$.
This along with (\ref{eqn_1231}) then proves (\ref{kprime_eqiv}).
Condition (D.3) of (\ref{theta_cnds_2}) is equivalent to
{
\vspace{-2mm}
\begin{eqnarray}
\label{eqn_gr}
g(R) & > & {\Big (} 1 + \frac{\rho_d}{\rho_r} {\Big )} \, \sqrt{\frac{R \rho_r}{\alpha}} \, =  \, g(x^{\prime}(R, \Theta))
\end{eqnarray}
}
and since $g(x)$ is strictly monotonically increasing with $x > 0$ we equivalently get
{
\vspace{-2mm}
\begin{eqnarray}
\label{eq_193}
x^{\prime}(R , \Theta) & < & R \,\,\,,\,\,\, \mbox{or equiv,} \,\, \frac{R}{x^{\prime}(R , \Theta)} \, >  \, 1.
\end{eqnarray}
}
Since $K_{max}(\Theta) > 1$ (from (\ref{theta_cnds_1})), using (\ref{eq_193}) in (\ref{opt_M_opt_K}) we finally get (\ref{lemma_w2}).
Similarly, if $K_{csi}^{\prime}(R, \Theta) > 1$, then from (\ref{opt_M_opt_K}) we have
$R > x^{\prime}(R, \Theta)$ which then implies (\ref{eqn_gr}), which in turn implies
(D.3) of (\ref{theta_cnds_2}).
$\hfill\blacksquare$

\begin{mylemma}
\label{lemma_234}
Consider any $\Theta$ satisfying (\ref{theta_cnds_1}) and $R_1 < R_2 < R_{max}(\Theta)$
such that both $(R_1, \Theta)$ and $(R_2 , \Theta)$
satisfy condition (D.3) of (\ref{theta_cnds_2}).
It follows that
{
\vspace{-2mm}
\begin{eqnarray}
\zeta_{csi}^{\prime}(R_1 , \Theta) & < & \zeta_{csi}^{\prime}(R_2, \Theta).
\end{eqnarray}
}
\end{mylemma}

{\em Proof}:
We will show that the partial derivative of $\zeta_{csi}^{\prime}(R , \Theta)$ w.r.t. $R$ is positive for any $(R, \Theta)$ with $\Theta$ satisfying (\ref{theta_cnds_1}), $(R, \Theta)$ satisfying (D.3) of (\ref{theta_cnds_2}) and $R < R_{max}(\Theta)$.
Since $R < R_{max}(\Theta)$ and $(R, \Theta)$ satisfies (D.3) of (\ref{theta_cnds_2}), from Lemma \ref{lemma_918}
it follows that
{
\vspace{-3mm}
\begin{eqnarray}
\label{eqn_234_234}
1 & < \, K^{\prime}_{csi}(R, \Theta) \, < & K_{max}(\Theta).
\end{eqnarray}
}
Using (\ref{eqn_234_234}) in (\ref{opt_M_opt_K}) we get
{
\vspace{-2mm}
\begin{eqnarray}
\label{app_234}
K^{\prime}_{csi}(R , \Theta) & = & \frac{R}{x^{\prime}(R , \Theta)}.
\end{eqnarray}
}
Using (\ref{app_234}) in (\ref{opt_M_opt_K}) and (\ref{zf_csi_opt_MK}) along with the definition of $g(x)$ we get
{
\vspace{-2mm}
\begin{eqnarray}
\label{zeta_xprime}
\frac{1}{\zeta_{csi}^{\prime}(R , \Theta)} &  = &   (\rho_r + \rho_d) h(x^{\prime}) + \frac{\rho_s}{R} \nonumber \\
\mbox{where} \,\, h(x^{\prime}) & \Define &  \frac{1 }{x^{\prime}} \, {\Bigg (}  1 + \frac{2 (2^{x^{\prime}} - 1) }{ 2^{x^{\prime}}  x^{\prime} \log(2) - 2^{x^{\prime}} + 1}  {\Bigg )}
\end{eqnarray}
}
where $x^{\prime}$ is used as an abbreviation for $x^{\prime}(R , \Theta)$.
Further
{
\vspace{-2mm}
\begin{eqnarray}
\label{deriv_zeta}
\frac{\partial \, {\Big ( } 1/\zeta_{csi}^{\prime}(R , \Theta) {\Big ) }}{\partial R} & \hspace{-2mm} = &  \hspace{-2mm} (\rho_r + \rho_d)
\frac{\partial h(x^{\prime}(R, \Theta))}{\partial R} \, - \,  \frac{\rho_s}{R^2}  \nonumber \\
& \hspace{-58mm} = & \hspace{-30mm}  (\rho_r + \rho_d) \, \frac{d h(x)}{d x} \vert_{x = x^{\prime}(R, \Theta)} \, \frac{\partial x^{\prime}(R, \Theta)}{\partial R} \, - \, \frac{\rho_s}{R^2}.
\end{eqnarray}
}
From Theorem \ref{thm1_stmt} we know that $g(x^{\prime}(R, \Theta)) = (1 + (\rho_d/\rho_r)) \sqrt{R \rho_r / \alpha}$ and therefore for a fixed $\Theta$,
$g(x^{\prime}(R, \Theta))$ increases strictly monotonically with increasing $R$.
Since $g(\cdot)$ is a strictly monotonically increasing function, it follows that $x^{\prime}(R, \Theta)$ increases strictly monotonically with increasing $R$, that is
${\partial x^{\prime}(R , \Theta)}/{\partial R}  >  0$.
Using this in (\ref{deriv_zeta}) along with the fact that $dh(x)/dx < 0$ we finally get
${\partial \, {\Big ( } 1/\zeta_{csi}^{\prime}(R , \Theta) {\Big ) }}/{\partial R} \, < \, 0$.
$\hfill\blacksquare$

\subsection{Lower bound on $\zeta_{zf}^{\prime}(\Theta)$}
\label{appendix_pw_1}
\begin{mylemma}
\label{lemma_lb}
For any $(R, \Theta)$ satisfying (\ref{theta_cnds_1}) and (\ref{theta_cnds_2}) we have
{
\vspace{-3mm}
\begin{eqnarray}
\label{lemma_lb_eqn}
\zeta_{zf}^{\prime}(R , \Theta) & > & \frac{3}{8} \, \zeta_{csi}^{\prime}{\Big (} \frac{4 R}{3} , \Theta {\Big )} 
\, >  \,  \frac{3}{8} \, \zeta_{csi}^{\prime}{\Big (} R , \Theta {\Big )}
\end{eqnarray}
}
\end{mylemma}

{\em Proof}:
The lower bound on $\zeta_{zf}^{\prime}(R , \Theta)$ follows from the following chain of
inequalities
{
\small
\begin{eqnarray}
\label{lb_seq}
\frac{R }{\zeta_{zf}^{\prime}(R , \Theta)} & \hspace{-2mm} = & \hspace{-6mm} \min_{\substack{ (M, K, \tau) \in {\mathbb R}^3 \, \vert \, \\
1 \leq K \leq  K_{max}(\Theta), \\
K \leq \tau < T \,,\,
M > K}}   {\Bigg (} \alpha K \gamma_u \, + \, K {\Big (} \rho_d \, + \, \frac{8}{3} K^2  \frac{\rho_0}{T} {\Big )}   \nonumber \\
& &  + \, \rho_s + M {\Big (} \rho_r \, + \, 2 K \rho_0 \, + \, 4 K^2 \frac{\rho_0}{T}{\Big )} {\Bigg )} \nonumber \\
& \hspace{-30mm} {(a) \atop < } & \hspace{-19mm} \min_{\substack{ (M, K, \tau) \in {\mathbb R}^3 \, \vert \, \\
1 \leq K \leq  K_{max}(\Theta) \\
\tau = K_{max}(\Theta) , M > K}}   \,  {\Bigg [} \, \frac{\alpha K}{(M - K)} {\Big (} 1 + \frac{K}{\tau} {\Big )} {\Big (}  2^{\frac{R}{{ K} ( 1 - \frac{{ \tau}}{T})}} - 1 {\Big )} \nonumber \\
 & &  \hspace{5mm} + \, \frac{\alpha K }{K + \tau}  \, + \, \rho_s \, + \, 2 M \rho_r \, + \, 2 K \rho_d {\Bigg ]} \nonumber \\
& \hspace{-30mm}  {(b) \atop < } & \hspace{-19mm}  \min_{\substack{ (M, K) \in {\mathbb R}^2 \, \vert \, \\
1 \leq K \leq  K_{max}(\Theta), \\
M > K}} \, {\Bigg [} 2 \frac{\alpha K}{(M - K)} {\Big (}  2^{\frac{R}{{ K} { (} 1 - ({K_{max}(\Theta)}/{T}) { )} }} - 1 {\Big )}   \nonumber \\
 & &  \hspace{5mm} + \,  2 \rho_s + 2 M \rho_r + 2 K \rho_d  {\Bigg ]} \nonumber \\
& \hspace{-30mm}  {(c) \atop \leq } & \hspace{-19mm}  \min_{\substack{ (M, K) \in {\mathbb R}^2 \, \vert \, \\
1 \leq K \leq  K_{max}(\Theta) \\
, M > K}} \hspace{-2mm} \frac{2 \alpha K {\Big (}  2^{\frac{4R}{3 K}} - 1 {\Big )}}{(M - K)} +   2 \rho_s + 2 M \rho_r + 2 K \rho_d    \nonumber \\
& {(d) \atop = }  & \, \frac{8R/3}{\zeta_{csi}^{\prime}( \frac{4R}{3} \,,\, \Theta)}
\,\,\, {(e) \atop < } \,\,\,  \frac{8R/3}{\zeta_{csi}^{\prime}( R \,,\, \Theta)}
\end{eqnarray}
\normalsize
}
where step (a) follows from Lemma \ref{lemma_gammau_ub}, Lemma \ref{lemma_theta_cnds} (condition (\ref{theta_cnds}) in Lemma \ref{lemma_theta_cnds} is implied by (\ref{theta_cnds_1}) and (D.1) of (\ref{theta_cnds_2})).
We have also used the fact that $\tau = K_{max}(\Theta)$ is a valid choice
since $\tau = K_{max}(\Theta) < T$ and $\tau = K_{max}(\Theta) \geq K$.
In step (b) we have used the fact that $(1 + (K/\tau)) = (1 + K/K_{max}(\Theta)) \leq 2$ since $K \leq  K_{max}(\Theta)$. In step (b) we have also used the fact that $\alpha K / (K + \tau)  \, + \, \rho_s < 2 \rho_s$ since $\alpha K / (K + \tau) \leq \alpha/2$ (as $K \leq  \tau$) and $\alpha/2 < \rho_s$ (from (D.1) of (\ref{theta_cnds_2})).{\footnote{\footnotesize{{Since for any $v > 0$, $\sqrt{1 + v} > 1$, from (\ref{eqn_19_thm})
in the proof of Lemma \ref{lemma_gammau_ub} it follows that $\gamma_u$ is lower bounded by
$\frac{({K} + { \tau}) \, {\Big (}  2^{\frac{R}{{ K} ( 1 - \frac{{ \tau}}{T})}} - 1 {\Big )}}{ { \tau} ({ M} - { K})}$.
Therefore, tightening of Lemma \ref{lemma_gammau_ub} can at best replace the term $1 / (K + \tau)$ in the R.H.S. of (\ref{eqn_22})
by a smaller positive value, say $\epsilon/ (K + \tau)$ for some $\epsilon < 1$.
Using this tightened bound of Lemma \ref{lemma_gammau_ub} in step (a) of (\ref{lb_seq}), the main result in
(\ref{lemma_lb_eqn}) will continue to hold even if $\rho_s > \epsilon \, \alpha/2$.
Since $\epsilon < 1$ it follows that tightening
of the bound in Lemma \ref{lemma_gammau_ub} will result in the condition $\rho_s > \alpha/2$
getting relaxed to $\rho_s > \epsilon \alpha/2$.
}
}}}
Step (c) follows from the fact that $K_{max}(\Theta) \leq T/4$ (see (\ref{kmax_def})).
Step (d) follows from (\ref{eqn_332}) and (\ref{zf_csi_opt_MK}).

Since $(R,\Theta)$ satisfies (\ref{theta_cnds_1}) and (\ref{theta_cnds_2}), from Lemma \ref{lemma_37}
it follows that $R_1 \Define R < 3R_{max}(\Theta)/4$ and therefore $R_2 \Define 4R/3 < R_{max}(\Theta)$.
Note that since $(R,\Theta)$ satisfies (D.3) of (\ref{theta_cnds_2}), it is clear that $(R_1=R,\Theta)$
also satisfies condition (D.3) of (\ref{theta_cnds_2}). Since $g(x)/\sqrt{x}$ is a strictly monotonically
increasing function with $x > 0$ and $R_2 > R_1$, it follows that $g^2(R_2)/R_2 > g^2(R_1)/R_1$.
Using this in (D.3) of (\ref{theta_cnds_2}) it is clear that $(R_2,\Theta)$ also satisfies this condition.
Step (e) now follows from Lemma \ref{lemma_234}.
$\hfill\blacksquare$
\subsection{Upper bound on $\zeta_{zf}^{\prime}(R, \Theta)$}
\label{appendix_pw_2}
\begin{mylemma}
\label{lemma_ub}
For $\Theta$ satisfying (\ref{theta_cnds_1}) and any $R > 0$ we have
{
\vspace{-2mm}
\begin{eqnarray}
\label{lemma_ub_eqn}
\zeta_{zf}^{\prime}(R , \Theta) & < & \zeta_{csi}^{\prime}{\Big (} R , \Theta {\Big )}.
\end{eqnarray}
}
\end{mylemma}

{\em Proof}:
An upper bound can be derived by considering $p_{mud} = 0$, i.e., from (\ref{prob_prime}) we have 
{
\small
\vspace{-2mm}
\begin{eqnarray}
\label{prob_prime_ub}
\frac{R }{\zeta_{zf}^{\prime}(R , \Theta)} & > & \hspace{-3mm} \min_{\substack{ (M, K, \tau) \in {\mathbb R}^3 \, \vert \, \\
1 \leq K \leq  K_{max}(\Theta), \\
K \leq \tau < T \,,\,
M > K}}  \alpha K \gamma_u \, + \, \rho_s \, + \, M \rho_r \, + \, K \rho_d \nonumber \\
& \hspace{-27mm} {(a) \atop > }  &  \hspace{-16mm} \min_{\substack{ (M, K) \in {\mathbb R}^2 \, \vert \, \\
1 \leq K  \leq K_{max}(\Theta) \\
M > K}}   \alpha K  \frac{(2^{R/K} - 1)}{(M - K)}\, + \, \rho_s \, + \, M \rho_r \, + \, K \rho_d  \nonumber \\
& =  & \frac{R}{\zeta_{csi}^{\prime}(R , \Theta)}
\end{eqnarray}
\normalsize
}
\normalsize
where the last equality follows from (\ref{eqn_332}) and (\ref{zf_csi_opt_MK}). Step (a) follows from the fact that $\gamma_u  >  \frac{K + \tau}{ \tau }  \, \frac{{\Big (} 2^{\frac{R}{K(1 - \frac{\tau}{T})} } - 1 {\Big )} }{(M - K)}  \,
>  \,  \frac{{\Big (} 2^{\frac{R}{K} } - 1 {\Big )} }{(M - K)}$ (see (\ref{ZF_sum_rate_gammau})).
$\hfill\blacksquare$

\subsection{Proof of Theorem \ref{decGc_deceta}}
\label{prf_decGc_deceta}
Since $\rho_r = G_c p_r / (N_0 B)$, it follows that (\ref{cnd_23}) is equivalent
to (\ref{theta_cnds_2}).
Since $(R, \Theta)$ satisfies (\ref{theta_cnds_1}) and (\ref{theta_cnds_2}), from
Lemma \ref{lemma_37} we have $R < R_{max}(\Theta)$ and therefore from Lemma \ref{lemma_918} we have $K^{\prime}_{csi}(R, \Theta) = R/x^{\prime}(R, \Theta)$
which gives us (\ref{zeta_xprime}).
From (\ref{zeta_xprime}) it follows that $\eta_{csi}^{\prime}(R , \Theta)$ satisfies
{
\vspace{-1mm}
\small
\begin{eqnarray}
\label{pdeqn}
\frac{(R B) / p_r }{\eta_{csi}^{\prime}(R , \Theta)} & \hspace{-3mm} = & \hspace{-4mm}  {\Bigg (} \frac{(1 + \beta) R}{x^{\prime}} {\Big (}  1 + \frac{2 (2^{x^{\prime}} - 1) }{ 2^{x^{\prime}}  x^{\prime} \log(2) - 2^{x^{\prime}} + 1}  {\Big )}  +  \delta{\Bigg )}  \nonumber \\
\end{eqnarray}
\normalsize
}
where $x^{\prime}$ is an abbreviation for $x^{\prime}(R , \Theta)$.
It can be shown that the partial derivative of the R.H.S. above w.r.t. $x^{\prime}(R, \Theta)$ is
negative and therefore since $p_r$ is constant we have
{
\vspace{-2mm}
\small
\begin{eqnarray}
\label{eqn_994}
\frac{\partial 1/\eta_{csi}^{\prime}(R , \Theta) }{\partial x^{\prime}(R, \Theta)} & < & 0.
\end{eqnarray}
\normalsize
}
Since $(\alpha, R, T, p_r, p_d, p_s, C_0)$ are fixed, $(\alpha, R, T, \beta, \delta, \mu)$ are also
fixed, and therefore from Lemma \ref{lemma_654} we also know that the partial derivative of $x^{\prime}(R , \Theta)$ w.r.t.
$\rho_r$ is positive and therefore since $p_r$ is fixed we have
$\frac{\partial x^{\prime}(R, \Theta)}{\partial G_c}  \, = \, \frac{p_r}{N_0 B} \, \frac{\partial x^{\prime}(R, \Theta)}{\partial \rho_r}  \, >  \, 0$.
Using this fact along with (\ref{eqn_994}) completes the proof.
$\hfill\blacksquare$

\subsection{Proof of Theorem \ref{decrhor_incMK}}
\label{prf_decrhor_incMK}
Since $(R, \Theta)$ satisfies the conditions in (\ref{theta_cnds_1}) and (\ref{theta_cnds_2}),
from Lemma \ref{lemma_918} it follows that
$K^{\prime}_{csi}(R , \Theta) = R/x^{\prime}(R , \Theta)$,
From Lemma \ref{lemma_654} we know that $x^{\prime}(R , \Theta)$
decreases with decreasing $\rho_r$, and since $K^{\prime}_{csi}(R , \Theta) = R/x^{\prime}(R , \Theta)$ it follows that $K^{\prime}_{csi}(R , \Theta)$
increases strictly monotonically with decreasing $\rho_r$.

Using $K^{\prime}_{csi}(R , \Theta) = R/x^{\prime}(R , \Theta)$ in (\ref{opt_M_opt_K})
along with the expression for $g(x)$, gives us
{
\vspace{-2mm}
\small
\begin{eqnarray}
\label{eqn_221}
M^{\prime}_{csi}(R , \Theta) & \hspace{-2mm} = & \hspace{-2mm} K^{\prime}_{csi}(R , \Theta) \, {\Big (}  1 + \frac{ (1 + \beta) (2^{x^{\prime}} - 1) }{ 2^{x^{\prime}}  x^{\prime} \log(2) - 2^{x^{\prime}} + 1}  {\Big )}
\end{eqnarray}
\normalsize
}
where $x^{\prime}$ is an abbreviation for $x^{\prime}(R, \Theta)$.
It can be shown that $s(x^{\prime}) \Define (2^{x^{\prime}} - 1)/(2^{x^{\prime}}  x^{\prime} \log(2) - 2^{x^{\prime}} + 1)$ increases strictly
monotonically with decreasing $x^{\prime} > 0$.
From Lemma \ref{lemma_654} we know that $x^{\prime}(R , \Theta)$ decreases strictly monotonically with decreasing $\rho_r$.
Combining these facts we see that $s(x^{\prime})$ increases strictly monotonically with decreasing $\rho_r$. Using this fact in (\ref{eqn_221}) along with the fact that $K^{\prime}_{csi}(R , \Theta)$ increases with decreasing $\rho_r$ it follows that $M^{\prime}_{csi}(R , \Theta)$ increases strictly monotonically
with decreasing $\rho_r$.
$\hfill\blacksquare$

\subsection{Proof of Theorem \ref{thm_fixK}}
\label{prf_thm_fixK}
In (\ref{bnd_eff_eqn}), the lower bound on the ratio ${\zeta_{zf}^{\prime}(R, \Theta)}/{\zeta_{zf}^{\prime}(K, R, \Theta)}$ follows from the fact that $K$ is not necessarily equal to the optimal $K^{\prime}_{zf}(R, \Theta)$.
Starting from the definition of $\zeta_{zf}^{\prime}(K, R, \Theta)$  in (\ref{eqn_fixK}) and using $k^{\prime}$ as an abbreviation for $K^{\prime}_{csi}(R, \Theta)$ we have
{
\small
\begin{eqnarray}
\label{bnd_eff_eqn_1}
\frac{R }{\zeta_{zf}^{\prime}(K , R , \Theta)} &   = & \hspace{-4mm} \min_{\substack{ (M, \tau) \in {\mathbb R}^2 \, \vert \, \\
K \leq \tau < T \,,\,
M > K}}   {\Bigg [} \alpha K \gamma_u \, + \, \rho_s   \nonumber \\ 
& &  \hspace{9mm}  + \, M {\Big (} \rho_r \, + \, 2 K \rho_0 \, + \, 4 K^2 \frac{\rho_0}{T}{\Big )}  \nonumber \\
& &  \hspace{9mm} + \, K {\Big (} \rho_d \, + \, \frac{8}{3} K^2  \frac{\rho_0}{T} {\Big )}  {\Bigg ]}  \nonumber \\
& \hspace{-45mm}  {(a) \atop < } & \hspace{-22mm}  \min_{\substack{ M \in {\mathbb R} \, \vert \, \\
, M > K}} \, {\Bigg [} 2 \frac{\alpha K}{(M - K)} {\Big (}  2^{\frac{4R}{3 K}} - 1 {\Big )}
\, + \,  2 \rho_s + 2 M \rho_r + 2 K \rho_d  {\Bigg ]}   \nonumber \\
&  \hspace{-45mm} {(b) \atop = } &  \hspace{-22mm}  2 {\Big [} K ( \rho_d + \rho_r)  + 2 \sqrt{\alpha \rho_r K  (2^{4R/3K} - 1) } \, + \, \rho_s {\Big ]}\nonumber \\
& \hspace{-45mm} { (c) \atop < } & \hspace{-20mm} 2 \frac{K}{K^{\prime}} \, {\Big [} K^{\prime} (\rho_d + \rho_r)
\, + \,  2  \sqrt{\alpha \rho_r  K^{\prime}  (2^{\frac{R}{K^{\prime}}} - 1) } \, + \, \rho_s  {\Big ]}    \nonumber \\
& \hspace{-45mm}  {(d) \atop = } &  \hspace{-22mm}  2 \, \frac{K}{K^{\prime}} \, \frac{R}{\zeta_{csi}^{\prime}(R \,,\, \Theta)} 
\end{eqnarray}
\normalsize
}
where step (a) is exactly similar to the sequence of steps (a), (b) and (c) in (\ref{lb_seq}) for a fixed $K$
(note that $\Theta$ satisfying (\ref{theta_cnds_1}) and $K  \leq  K_{max}(\Theta)$ are sufficient conditions for steps (a), (b), and (c) in (\ref{lb_seq}) to hold).
Step (b) follows from (\ref{opt_M_given_K}) and (\ref{zeta_zf_K}) for a sum-rate of $4R/3$.
Step (c) follows from the fact that $K > 4K_{csi}^{\prime}(R, \Theta)/3$.
Step (d) follows from Theorem \ref{thm1_stmt} (in the proof of Theorem \ref{thm1_stmt},
substituting $K=K_{csi}^{\prime}(R, \Theta)$ in (\ref{zeta_zf_K}) gives us $1/\zeta^{\prime}_{csi}(R, \Theta)$).
Further from Lemma \ref{lemma_ub} it follows that
$\zeta_{zf}^{\prime}(R, \Theta)  <  \zeta_{csi}^{\prime}(R, \Theta)$.
Using this along with (\ref{bnd_eff_eqn_1}) we get
the upper bound on $\zeta_{zf}^{\prime}(R, \Theta)/\zeta_{zf}^{\prime}(K, R, \Theta)$
in (\ref{bnd_eff_eqn}).
$\hfill\blacksquare$

\begin{IEEEbiography}
{Saif Khan Mohammed} (S'08-M'11) Saif Khan Mohammed (S'08-M'11) received the B.Tech degree in
Computer Science and Engineering from the Indian Institute of Technology (I.I.T.),
New Delhi, India, in 1998 and the Ph.D. degree from the Electrical and Communication
Engineering Department, Indian Institute of Science, Bangalore, India, in 2010.
Currently he is an Assistant Professor in the Department of Electrical Engineering, I.I.T. Delhi.
From Sept. 2011 to Feb. 2013, he was Assistant Professor at the Communication Systems Division (Commsys) in the Electrical Engineering Department (ISY) at Linkoping University, Sweden.
From 2010 to 2011, he was a postdoctoral researcher at Commsys.
He has previously worked as a Systems and Algorithm designer in the Wireless Systems Group at Texas Instruments, Bangalore (India) (2003 - 2007).

From 2000 to 2003, he worked with Ishoni Networks, Inc., Santa Clara, CA (USA), as a Senior Chip Architecture Engineer.
From 1998 to 2000, he was a ASIC Design Engineer with Philips, Inc., Bangalore.
His main research interests include wireless communication
using large antenna arrays,
coding and signal processing for wireless communication systems, and statistical
signal processing.
He is a member of the IEEE, the IEEE Communication Society, the IEEE Signal Processing Society and the IEEE Information Theory Society. He has been a Technical Program Committee member for several IEEE sponsored conferences (International Conference on Communications (ICC' 2013, 2014), the IEEE Vehicular Technology Conference (VTC)  Spring 2013, and the IEEE Swedish Communication Theory Workshop (Swe-CTW) Fall 2012).
He has also served as a guest editor for a special issue on Massive MIMO in the Journal of Communication Networks (JCN). He also holds four US patents on detection and precoding of Massive MIMO signals.

Dr. Mohammed was awarded the Young Indian Researcher Fellowship by the Italian Ministry of University and Research (MIUR) for the year 2009-10.
He was also awarded the CENIIT (Linkoping University) research grant for the year 2012.
\end{IEEEbiography}

\end{document}